\documentclass[11pt]{article}
\usepackage{amsmath,amssymb}
\usepackage{graphicx}
\usepackage{booktabs}
\usepackage{url}
\usepackage{enumitem}
\usepackage{authblk}
\usepackage{geometry}
\usepackage{natbib}
\usepackage{caption}
\usepackage{subcaption}

\title{Characterising Behavioural Families and Dynamics of Promotional Twitter
 Bots via Sequence-Based Modelling}

\author[1]{Ohoud Alzahrani}
\affil[1]{Department of Computer Science, Umm Al-Qura University, Saudi Arabia}
\author[2]{Russell Beale}
\author[2]{Robert J. Hendley}
\affil[2]{School of Computer Science, University of Birmingham, UK}
\date{}

\begin{document}
\maketitle

\begin{abstract}
This paper investigates whether promotional Twitter bots can be meaningfully
grouped into behavioural families and whether members of a given family exhibit
similar behavioural evolution over time. We analyse 2{,}798{,}672 tweets from
2{,}615 ground-truth promotional bot accounts active on Twitter between 2006 and
2021, restricting our analyses to complete calendar years from 2009 to 2020. We
encode each bot's activity as a sequence of symbolic blocks - a ``digital DNA''
representation - derived from seven categorical post-level behavioural features
(posting action, URL, media, text duplication, hashtags, emojis, and sentiment),
with the post timestamp retained only to preserve temporal ordering. Using
non-overlapping blocks (equivalently $k$-mers with $k=7$), we compute pairwise
cosine similarities over block-frequency vectors, construct a similarity
matrix, and apply hierarchical clustering to obtain bot families. We then
characterise each family in terms of dominant feature usage, shared and unique
blocks, and temporal changes across three life-cycle segments (beginning,
middle, end). The results show that promotional bots can be clustered into four
coherent families - \emph{Unique Tweeters}, \emph{Duplicators with URLs},
\emph{Content Multipliers}, and \emph{Informed Contributors} - that share common
behavioural cores but differ systematically in engagement strategies and
behavioural evolution.

Building on this family structure, we introduce a sequence-analysis framework
for modelling bot behaviour change explicitly as behavioural mutations. Within
each family, we align behavioural sequences using multiple sequence alignment
(MSA), then detect and classify changes as insertions, deletions, substitutions,
alterations, and identity (match) events. This allows us to quantify mutation
frequencies, identify blocks most prone to change, analyse which behavioural
features are most often substituted, and locate mutation hotspots along the
sequences. The analysis shows that deletions and substitutions dominate while
insertions are rare; that mutation patterns differ systematically across the
four families; and that mutation hotspots concentrate in early sequence
positions for some families while being widely distributed for others.

Finally, we evaluate whether family structure and mutation patterns support
predictive reasoning about behavioural adaptation. We show that bots within the
same family are more likely to share mutations than bots in different families,
that closely related bots within a family share and propagate mutations more
often than distant ones, and that behaviour around external trigger events (such
as Christmas and Halloween) follows family-specific, partially predictable
patterns. These findings support sequence-based family modelling of promotional
bots together with a fine-grained, evolution-inspired account of how their
behaviours adapt and evolve over time.
\end{abstract}

\section{Introduction}
Automated accounts are a persistent feature of major social-media platforms. Among them, \emph{promotional bots}—accounts that systematically advertise products, services, events, or content—are particularly important because they operate at scale and can distort attention and engagement metrics, amplify marketing campaigns, and contribute to spam-like ecosystem effects. While many studies focus on \emph{detecting} bots at a given point in time, the practical challenge is increasingly longitudinal: bot operators adapt to platform changes and defensive measures, and so the behaviours that define a bot today may not be the same behaviours that define it tomorrow.

\paragraph{Overview.}
We analyse longitudinal behaviour of promotional Twitter\footnote{Twitter was rebranded as X in 2023; we use ``Twitter'' to reflect the period covered by the dataset.} bots using 2{,}615 ground-truth accounts drawn from publicly available research corpora (2{,}798{,}672 tweets). Focusing on complete calendar years (2009--2020), we represent each account as an ordered sequence of compact behavioural blocks derived from seven categorical posting features. We then (i) cluster accounts into coherent behavioural families and (ii) characterise within-family change using sequence alignment and a small set of mutation operators (insertion, deletion, substitution, alteration, and identity/no-change events). The result is an interpretable, end-to-end pipeline for large-scale behavioural trace analysis that reveals not just whether bots can be detected, but how they adapt and which aspects of that adaptation are shared within families.

This motivates a shift from treating bot behaviour as a single static signature to treating it as a \emph{structured, evolving process}. Two questions become central. First, are there meaningful \emph{behavioural families} of promotional bots—coherent groupings in which bots share common behavioural “cores” but differ in systematic ways? Second, if such families exist, how do their behaviours \emph{change over time}, and can those changes be described in a way that supports both explanation and (at least limited) prediction? Addressing these questions matters not only for bot analysis, but also for the robustness of downstream detection systems: when behaviour drifts, feature distributions and feature-structure relationships that detectors rely on can become unreliable.

To study both family structure and temporal evolution, we model each bot’s activity stream as an ordered sequence of symbolic blocks (a \emph{digital DNA} representation). Each block encodes a compact combination of post-level behavioural and content attributes (e.g., posting action, URL/media use, text duplication, hashtags, emojis, sentiment, and coarse timing). This sequence view allows us to compare bots at the level of \emph{behavioural pattern distributions} rather than isolated features. We first estimate bot-to-bot similarity using non-overlapping $k$-mers derived from these sequences and cosine similarity over the resulting pattern vectors, then apply hierarchical clustering to obtain a behavioural “family tree”. We characterise each family by its dominant feature usage, its common and unique blocks, and how its dominant patterns shift across life-cycle segments (beginning, middle, end).

We then extend this family analysis with a second, bioinformatics-inspired stage that makes behaviour change explicit. Within each family, we align behavioural sequences using multiple sequence alignment (MSA) and treat differences between aligned sequences as \emph{behavioural mutations}. We categorise changes into insertions, deletions, substitutions, alterations, and recurrent “identity” events. This enables quantitative answers to questions such as: which blocks (and which underlying behavioural features) are most prone to change; where along the activity sequence change concentrates (mutation “hotspots”); and how mutation profiles differ across families. Finally, we assess whether this evolutionary framing has predictive value, asking whether bots within the same family (and especially close relatives) tend to share and propagate mutations, and whether families exhibit partially repeatable responses around external trigger events (e.g., seasonal holidays).

\paragraph{Research questions.}
The work reported in this paper addresses four closely related questions about the behaviour of promotional Twitter bots:

\noindent\textbf{RQ1} Is there evidence that the behaviour of promotional bots can be classified into distinct families,
such that members of each family exhibit similar behaviour?

\noindent\textbf{RQ2} Do promotional bots within the same family exhibit similar changes in behaviour over time?

\noindent\textbf{RQ3} Can we model bots’ behaviour change in terms of adding, removing, or modifying activities?

\noindent\textbf{RQ4} Is the metaphor of genetic evolution useful for the prediction of upcoming changes in bot
behaviour?

\paragraph{Contributions}
We make five contributions aimed at reproducible analysis of large-scale digital traces:
\begin{itemize}
    \item a compact, interpretable \emph{sequence representation} of multi-feature
    posting behaviour (seven categorical features with timestamps retained for
    ordering) suitable for large-scale comparison across accounts;
    \item an end-to-end \emph{family discovery} pipeline (block-frequency
    vectors, cosine similarity, hierarchical clustering) that yields coherent
    behavioural families spanning multiple platform generations;
    \item a \emph{within-family sequence-analysis framework} that operationalises
    behavioural change via alignment and a small set of mutation operators
    (insertion, deletion, substitution, alteration, and identity/no-change events), enabling fine-grained
    localisation and quantification of change;
    \item empirical evidence that behavioural change is \emph{structured}:
    families share stable cores yet differ systematically in life-cycle dynamics,
    mutation hotspots, and feature substitutions, and these regularities support
    predictive reasoning about adaptation under shared pressures; and
    \item a clearly specified, implementation-ready methodology that supports
    reproducibility via shareable derived artefacts (feature sequences, block
    vocabularies, similarity matrices, and aggregate mutation statistics) even
    when raw content sharing is constrained by platform terms.
\end{itemize}

\section{Related Work}
\label{sec:related}

Research on automated social media accounts (\emph{social bots}) has grown quickly over the last decade. Surveys and typologies document how bots participate in political manipulation, spam, scams and misinformation at scale, and they map the design space of bot types and detection techniques \citep{cresci_decade_2020,gorwa_unpacking_2020,orabi_detection_2020,wu_twitter_2018,ellaky_systematic_2023,aljabri_machine_2023}. A recurring theme is that bots can dominate activity around contentious topics, and that effective detection must keep pace with evolving, increasingly sophisticated adversaries.

From the standpoint of complex data on online systems, two recurring challenges are (i)
non-stationarity in behavioural traces (concept drift over time) and (ii) the need for
representations that support both scalable computation and human-interpretable auditing.

We organise the discussion around (i) detection work on Twitter, (ii) sequence-based representations of behaviour, (iii) temporal dynamics and adaptation, and (iv) campaign- and coordination-oriented automation, before positioning the present work.

\subsection{Social bots and detection on Twitter}

A large body of work models bots at the account level, contrasting their
properties with those of genuine users. \citet{varol_online_2017} propose a
supervised learning framework that distinguishes humans from bots on Twitter
using a rich feature set drawn from account metadata, content statistics, and
network features, and show that hybrid feature families are required to achieve
high accuracy. \citet{gilani_large-scale_2019} conduct a large-scale
behavioural analysis of bots and humans on Twitter and demonstrate systematic
differences in how the two groups use mentions, hashtags, URLs and media, as
well as in their relative impact on information flows. 

More recently, large, carefully curated graph-based benchmarks have become
central to evaluation. TwiBot-22 provides a multi-relational Twitter graph with
high-quality manual labels and an associated evaluation framework that
re-implements a wide range of baselines, enabling systematic comparison of
graph-based detectors \citep{feng_twibot22_2022}. Building on such
datasets, \citet{dehghan_detecting_2023} study graph-structural and
node-embedding features for bot detection on TwiBot-20 and an Italian election
network, showing that structural embeddings capture discriminative patterns
beyond profile and content features and that hybrid feature sets combining
multiple embeddings can improve predictive power.

Many subsequent studies refine feature sets and modelling techniques. \citet
{loyola-gonzalez_contrast_2019} introduce contrast pattern-based classification
for bot detection, using discriminative behavioural patterns to separate bots
from humans. \citet{heidari_empirical_2021} compare a range of machine learning
algorithms for social media bot detection and report that performance depends
strongly on feature engineering and dataset characteristics . Nicola et al.\
revisit classic feature sets and show that ``old'' features retain significant
predictive power for newer generations of bots, but also highlight performance
degradation when facing previously unseen bot classes \citep
{nicola_efficacy_2021}. Rovito et al.\ explore an evolutionary computation
approach for Twitter bot detection, tuning detection models via evolutionary
optimisation in order to adapt to changing bot behaviour \citep
{rovito_evolutionary_2022}.

Alongside structural and temporal features, several works exploit sentiment and
content-based cues. Dickerson et al.\ use sentiment features to distinguish
bots from humans and ask whether humans are systematically more opinionated
than bots in political conversations \citep{dickerson_using_2014}. Monica and
Nagarathna propose a sentiment-analysis-based method for detecting fake
tweets \citep{monica_detection_2020}, while Phan et al.\ show that
feature-ensemble models can improve sentiment analysis performance on tweets
with fuzzy sentiment, which can in turn support more effective content-based
detection pipelines \citep{phan_improving_2020}. 

Recent work has also stressed the adversarial nature of bot detection. Cresci
argues that the field is entering an era of adversarial social bot detection,
where bots are explicitly adapted to evade existing detectors and where
robustness and generalisation are as important as raw accuracy \citep
{cresci_coming_2021}. Cresci's ten-year overview of social bot detection
similarly traces an arms race between detection systems and increasingly
sophisticated bots \citep{cresci_decade_2020}. Systematic reviews of detection
systems \citep
{orabi_detection_2020,ellaky_systematic_2023,aljabri_machine_2023} and spam
detection surveys \citep{wu_twitter_2018} reinforce this view, highlighting
issues of reproducibility, dataset bias, and the difficulty of maintaining
performance as bots evolve. Complementing these perspectives, Giroux et al.\
argue that progress in detection must be assessed not only through point
estimates but also through calibrated uncertainty, proposing a Bayesian neural
network detector that quantifies uncertainty over bot labels and allows trading
coverage for confidence in practical decision making \citep
{giroux_unmasking_2024}.

\subsection{Digital-DNA-inspired and sequence-based behavioural models}

Within this broader landscape, the idea of using sequence-based representations
to capture online behaviour has gained traction. Rather than representing
accounts as unordered feature vectors, these approaches encode activity
histories as symbolic sequences and apply tools from information theory and
bioinformatics. Gilmary et al.\ introduce a DNA-influenced approach in which
Twitter accounts are represented via digital-DNA-like sequences and relative
entropy over these sequences is used to detect automated behaviour \citep
{gilmary_dna-influenced_2022}. Their results show that entropy-based features
derived from activity sequences can discriminate bots from humans and help
surface coordinated automation, while abstracting away from specific textual
content.

Sequence-oriented methods are often complemented by techniques for uncovering
coordination. Pacheco et al.\ propose a framework for uncovering coordinated
networks on social media, combining similarity in posting behaviour
(including temporal and content similarity) with network analysis to detect
groups of accounts acting in concert \citep{pacheco_uncovering_2021}. Although
their focus is on coordination rather than per-account classification, the
underlying assumption is similar: that accounts can be characterised by
structured traces of behaviour which, when compared, reveal common scripts and
campaign routines.

Complementary to symbolic sequence modelling, Di Paolo et al.\ transform
accounts' temporal online behaviours into images using chaos-game
representations and then apply convolutional neural networks for bot detection,
showing that image-based models can capture local motifs in behavioural traces
that are difficult to hand-engineer as features \citep
{di-paolo_novel_2023}. Their work can be seen as an image-based analogue of
digital-DNA approaches, further reinforcing the value of rich sequential
encodings for automation detection.

These sequence-based, image-based and coordination-focused approaches offer two
advantages that are directly relevant to this work. First, they abstract away
from specific textual content or topical domains, making them more robust to
topic shifts and multilingual data. Second, they allow the reuse of mature
tools from bioinformatics, information theory, and computer vision - including
alignment, motif analysis, entropy-based similarity metrics and convolutional
architectures - for modelling behavioural regularities. However, existing work
typically treats sequences as static snapshots used for detection or
coordination discovery, rather than explicitly modelling \emph{how} these
sequences change over successive bot ``generations'' or within curated bot
families.

\subsection{Temporal dynamics and evolving spambots}

A complementary line of work investigates the temporal dynamics of user and bot
activity. Kooti et al.\ analyse user behaviour at the level of Twitter
sessions, showing that activity and engagement vary markedly across short-term
sessions and that such dynamics can be exploited to understand and predict user
engagement patterns \citep{kooti_twitter_2016}. Lee et al.\ conduct a
seven-month longitudinal study of ``content polluters'' on Twitter and
demonstrate that spam campaigns can be sustained through waves of account
creation and deletion, revealing long-term adaptation in response to platform
interventions \citep{lee_seven_2011}. Sedhai and Sun examine 14 million tweets
and characterise hashtag-oriented spamming, identifying large-scale patterns in
hashtag usage and proposing features for detecting spammy hashtags and
accounts \citep{sedhai_analysis_2017}.

Temporal studies have also explicitly contrasted human and bot dynamics. Pozzana
and Ferrara propose a framework for measuring bot and human behavioural
dynamics, modelling time-varying features and showing that dynamic patterns can
improve the separation between bots and humans beyond what static features
achieve \citep{pozzana_measuring_2020}. Torres-Lugo et al.\ analyse large-scale
tweet deletions on Twitter and show that deletions can be used strategically to
manipulate apparent behaviour and evade retrospective analyses and detection
systems, adding another temporal dimension to bot activity and clean-up \citep
{torres-lugo_manipulating_2022}. Work on coordinated networks similarly uses
temporal and content alignment as signals of orchestrated campaigns \citep
{pacheco_uncovering_2021}.

Experimental work that deploys bots and tracks their interaction patterns over
time also highlights the emergence of distinctive local structures. Alrhmoun
and Kertész create controlled ecosystems of Twitter bots and humans,
implementing several bot strategies and using network motifs to characterise
how different strategies shape local interaction patterns and how these
patterns evolve \citep{alrhmoun_emergent_2023}. Their findings reinforce the
view of bot ecosystems as self-organising and adaptive, but they do not follow
curated bot families across multiple engineered generations.

These studies collectively highlight that bot activity is dynamic and adaptive:
accounts change posting rhythms, content mix, and even their visible histories
over time. Nevertheless, temporal analyses are typically based on aggregate
counts (e.g., tweet volumes, retweet activity) or coarse temporal features,
rather than on symbolic behavioural sequences that can support detailed
mutation-style analyses of how specific behavioural motifs are introduced,
modified, or dropped.

\subsection{Bots, misinformation, and campaign-oriented automation}

Beyond generic spam and automation, bots are now widely recognised as important
actors in misinformation and campaign-oriented communication. Al-Rawi et al.\
analyse COVID-19-related tweets and show that bots act as active news
promoters, systematically amplifying certain outlets and narratives in pandemic
coverage \citep{al-rawi_bots_2020}. Zhang et al.\ study COVID-19 vaccine
discussions on Twitter and ask whether social bots' sentiment engagement shapes
public sentiment, finding that bots participate deeply in vaccine-related
discussions and that their sentiment patterns can influence the overall tone of
the conversation \citep{zhang_could_2022}. 

Other work examines domain-specific and demographic aspects of automation.
Puertas et al.\ use sociolinguistic features to perform gender profiling for
bots on Twitter, illustrating how language style can reveal both gender cues
and automated behaviour \citep{puertas_bots_2019}. Mouronte-L\'opez et al.\
analyse patterns of human and bot behaviour in Twitter conversations about
sustainability, showing that bots are tailored to particular topical domains
and that their interactions with human communities can subtly shape
discourse \citep{mouronte-lopez_patterns_2024}. As attention has shifted from
single accounts to campaign-level operations, coordination has become a key
concept. Cinelli et al.\ study coordinated inauthentic behaviour in retweet
cascades around the 2019 UK election, showing how coordinated accounts occupy
strategically central positions in cascades and quantifying their marginal
impact on information reach \citep{cinelli_coordinated_2022}. 

These domain-specific and coordination-focused studies complement generic
detection research by highlighting that bots are often deployed as part of
structured campaigns, with reusable templates, scripts and engagement
strategies tuned to particular issues or audiences. However, they typically
focus on characterising the deployed bots at a given point in time, rather than
reconstructing how successive generations of campaign bots adapt over the life
of a campaign or across related campaigns.

\subsection{Positioning the present work}

Prior work shows that social bot detection has matured into a diverse field of
methods, datasets and surveys, including large graph-based benchmarks and
embedding-based models that exploit structural features and explicitly model
uncertainty \citep
{cresci_decade_2020,orabi_detection_2020,wu_twitter_2018,ellaky_systematic_2023,aljabri_machine_2023,feng_twibot22_2022,dehghan_detecting_2023,giroux_unmasking_2024}.

It is demonstrated that sequence-based, digital-DNA-inspired and image-based
approaches provide powerful abstractions for modelling online behaviour and
automation \citep
{gilmary_dna-influenced_2022,di-paolo_novel_2023,pacheco_uncovering_2021},
whilst temporal, experimental and motif-based analyses reveal that bots adapt
their activity and local network structures over time \citep
{kooti_twitter_2016,lee_seven_2011,sedhai_analysis_2017,pozzana_measuring_2020,torres-lugo_manipulating_2022,alrhmoun_emergent_2023}/
Finally, campaign- and coordination-oriented studies document how bots are
embedded in specific topical and advocacy contexts and how coordinated accounts
influence information diffusion \citep
{al-rawi_bots_2020,zhang_could_2022,puertas_bots_2019,mouronte-lopez_patterns_2024,cinelli_coordinated_2022}.

Existing digital-DNA and sequence-based work, however, focuses primarily on
distinguishing bots from humans or on uncovering coordinated groups at a
particular moment \citep
{gilmary_dna-influenced_2022,cinelli_coordinated_2022}. Temporal studies tend
to work with aggregate activity counts rather than rich symbolic behavioural
sequences \citep{pozzana_measuring_2020}, and adversarial perspectives
emphasise the evolution of detection systems more than the detailed evolution
of bot behaviour \emph{within} curated families \citep
{cresci_coming_2021,nicola_efficacy_2021,rovito_evolutionary_2022,alrhmoun_emergent_2023,giroux_unmasking_2024}.
This research addresses these gaps by representing multiple generations of
promotional bots as behavioural sequences, clustering them into coherent
families, and applying sequence-analysis and mutation-inspired techniques to
reconstruct and quantify how their behaviour changes across generations and
over time.

From a data science perspective, the emphasis is on an interpretable,
end-to-end pipeline for extracting, representing, and analysing longitudinal
behavioural traces at scale. The sequence representation provides a compact
derived dataset that can be shared and re-used, and the family and mutation
analyses yield reusable summary artefacts (family assignments, block
distributions, and mutation statistics) that support reproducible comparison
across datasets, time periods, and platforms.

\section{Methodology}

\subsection{Data}

We focus on promotional spambots active on Twitter between 2006 and 2021. We
start from ground-truth bot IDs in publicly available research corpora,
including multiple datasets from the MIB project and Botometer's Bot
Repository. These datasets primarily comprise promotional and spammer accounts
and have been widely used in bot-detection research. Using the Twint Python
library, we collected publicly visible tweets for these bot accounts via the
platform's web interface (i.e., not via the official Twitter API). For each
account, we collected up to 3{,}200 of the most recent tweets, along with
associated metadata.

In total, we obtained 2{,}798{,}672 tweets from 2{,}615 unique bot accounts. We
restrict our analysis to the period 2009--2020 in order to work with complete
calendar years and to avoid gaps caused by collection interruptions.

\paragraph{Data access and ethics.}
We analyse only publicly available posts and metadata and do not attempt to infer
private information or attribute accounts to specific individuals. Because tweet
availability can change over time (e.g., deletion or suspension) and redistribution
of raw content may be constrained by platform terms, we centre the study on derived
behavioural features and symbolic sequences. This design supports replication on
tweet-ID based archives or other re-collectable platform snapshots.

Two aspects of the collection and representation are important for interpreting
results and for reproducibility. First, Twint provides at most 3{,}200 recent
tweets per account; consequently, for long-lived accounts the earliest activity
may be under-represented. Second, tweet availability can change over time (e.g.,
due to deletion or suspension). To reduce sensitivity to these effects, our
primary analyses use complete calendar years (2009--2020) and operate on
derived behavioural features and symbolic blocks rather than raw text content,
allowing the methodology to be replicated on tweet-ID based collections or
other platform archives.

For each bot $b_i$ in the set of bots $\textit{Bots} = \{bot_1, \ldots, bot_k\}$, we have a collection of posts:
\[ P(b_i) = \{P_1(b_i), P_2(b_i), \ldots, P_{n_i}(b_i)\}.
\]

From each post we extract seven categorical behavioural features and the post timestamp (used only to preserve ordering); let $F = \{f_1,\dots,f_8\}$ denote this set, where the final feature is the timestamp:
\begin{itemize}
    \item $f_1$: posting action (tweet, retweet, reply),
    \item $f_2$: URL presence,
    \item $f_3$: sentiment (positive/negative/neutral),
    \item $f_4$: media (image, video, none),
    \item $f_5$: text duplication (duplicate, unique, empty),
    \item $f_6$: hashtags (present/absent),
    \item $f_7$: emojis (present/absent),
    \item $f_8$: timestamp (date and time).
\end{itemize}

The seven categorical features $f_1$--$f_7$ are used for encoding; $f_8$ is
retained only for temporal ordering.

\subsection{Feature Extraction}

For each post $P_j(b_i)$ we extract categorical values for the seven
non-temporal features. Let $\hat{F} \subset F$ denote these features
(excluding the timestamp). Each feature $f \in \hat{F}$ has a finite set of
possible values $V_f$. The extracted feature vector for a post is
\begin{equation}
\hat{F}(P_j(b_i)) = \{v_{f_1}, v_{f_2}, \dots, v_{f_{|\hat{F}|}}\},\quad v_f \in
 V_f, \ \forall f \in \hat{F}.
\label{eq:feat}
\end{equation}

\subsection{Feature Encoding}

To support efficient comparison, we map each categorical feature value to a
unique alphabetic symbol using an encoder $E^{\text{feature}}_k : v \mapsto
l$. 

\begin{itemize}
    \item Posting action: T (tweet), R (retweet), Y (reply),
    \item URL: U (contains URL), X (no URL),
    \item Sentiment: P (positive), N (negative), L (neutral),
    \item Media: I (image), V (video), M (no media),
    \item Text: D (duplicate), Q (non-duplicate), E (empty),
    \item Hashtag: H (has hashtags), Z (no hashtags),
    \item Emoji: J (has emojis), K (no emojis).
\end{itemize}

The encoded representation of post $P_j(b_i)$ is an unordered set of letters:
\begin{equation} L(P_j(b_i)) = \{l_{f_1}, l_{f_2}, \dots, l_{f_{|\hat{F}|}}\},
\label{eq:encoded}
\end{equation} where each $l_f$ is the symbol assigned to $v_f$ via the
 encoder. 

\subsection{Constructing Behavioural Blocks}

Next, we construct an ordered, fixed-length block per post by arranging the
encoded symbols in a fixed feature order:
\begin{equation}
\text{block}_j = [l_{\text{PostingActions}},\, l_{\text{URLs}},\, l_{\text
 {Media}},\, l_{\text{Emoji}},\, l_{\text{Hashtag}},\, l_{\text{Text}},\, l_
 {\text{Sentiment}}].
\label{eq:block}
\end{equation} For a bot $b_i$, the set of all blocks is
\begin{equation}
\text{Blocks}(b_i) = [\text{block}_1, \text{block}_2, \dots, \text{block}_
 {n_i}],
\label{eq:blocks}
\end{equation} where each block corresponds to one post. For example, a simple
 post might be encoded as
\[
\text{block}_1 = [T\, X\, M\, K\, Z\, D\, L] = \text{TXMKZDL},
\] indicating a tweet without URL or media, no emoji or hashtag, duplicated
 text, and neutral sentiment.

\begin{figure}
    \centering
    \includegraphics[width=0.95\linewidth]{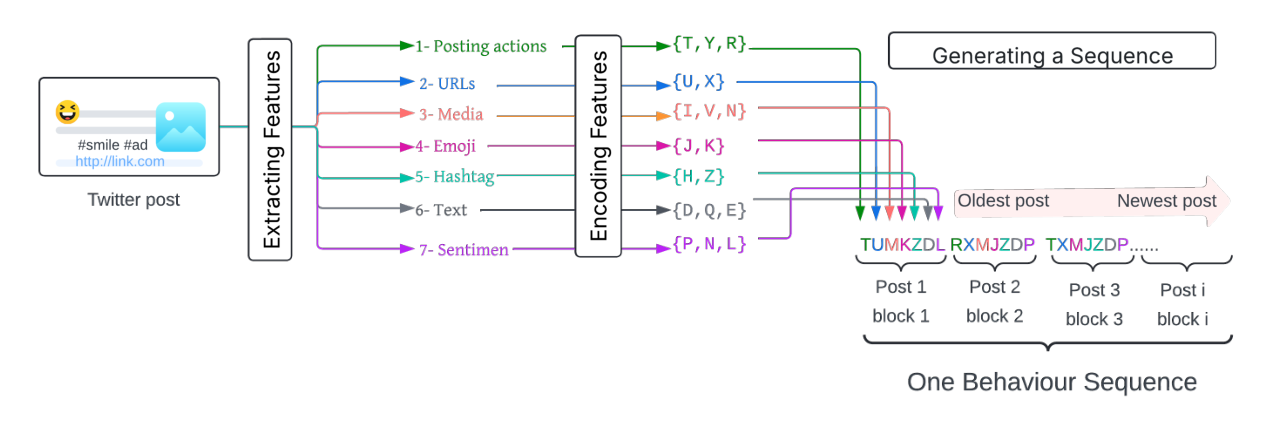}
    \caption{Process flow for feature extraction and encoding into behavioural
     sequences}
    \label{fig:5.2}
\end{figure} Figure \ref{fig:5.2} shows how feature extraction and encoding map
 into these behavioural blocks. 

\subsection{Building Behavioural Sequences}

To capture temporal evolution, we sort each bot's blocks by their associated
timestamps $t_j$ (extracted from $f_8$) in ascending order. We then concatenate
them to form a single behavioural sequence:
\begin{equation} S(b_i) = \text{Concatenate}(\text{Blocks}(b_i)) = [\text
 {block}_1 \ \text{block}_2 \ \dots \ \text{block}_{n_i}].
\label{eq:seq}
\end{equation} The length of the sequence is
\begin{equation}
\text{Len}(S(b_i)) = 7 \times n_i,
\end{equation} because each block has seven encoded letters. This sequence
 functions as the bot's ``digital DNA'' and supports subsequent similarity
 analysis. 

\subsection{Block-Frequency Vectors and Similarity Matrix}

To compare bots at the pattern level, we transform each behavioural sequence
into a block-frequency vector using non-overlapping units with $k=7$ (i.e., one
unit corresponds exactly to one block, since each block has length 7). For bot
$b_i$, the block-frequency vector is:
\begin{equation}
\text{Vec}_i = \{\text{Count}(\text{block}, S(b_i))\},
\label{eq:vec}
\end{equation} where $\text{Count}(\cdot)$ counts occurrences of each distinct
 block in $S(b_i)$. This representation preserves the interpretability of
 blocks while aggregating behaviour frequency over the entire timeline.

Given frequency vectors $\text{Vec}_i$ and $\text{Vec}_j$ for bots $b_i$ and
$b_j$, we compute cosine similarity:
\begin{equation} M[i,j] = \cos(\text{Vec}_i, \text{Vec}_j) = \frac{\text
 {Vec}_i \cdot \text{Vec}_j}{\|\text{Vec}_i\| \,\|\text{Vec}_j\|},
\label{eq:cosine}
\end{equation} producing an $m \times m$ similarity matrix $M$ for $m$ bots. The
 matrix is symmetric with $M[i,i]=1$.  

\subsection{Hierarchical Clustering into Families}

We convert similarities into dissimilarities via
\begin{equation} D[i,j] = 1 - M[i,j], \quad \forall i,j,
\label{eq:dissim}
\end{equation} and apply agglomerative hierarchical clustering with average
 linkage to the dissimilarity matrix $D$. This produces a dendrogram
 representing how bots merge into clusters at different dissimilarity levels.
 Each leaf on the $x$-axis corresponds to a bot; the height of each merge on
 the $y$-axis reflects inter-cluster dissimilarity.

\begin{figure}
    \centering
    \includegraphics[width=1\linewidth]{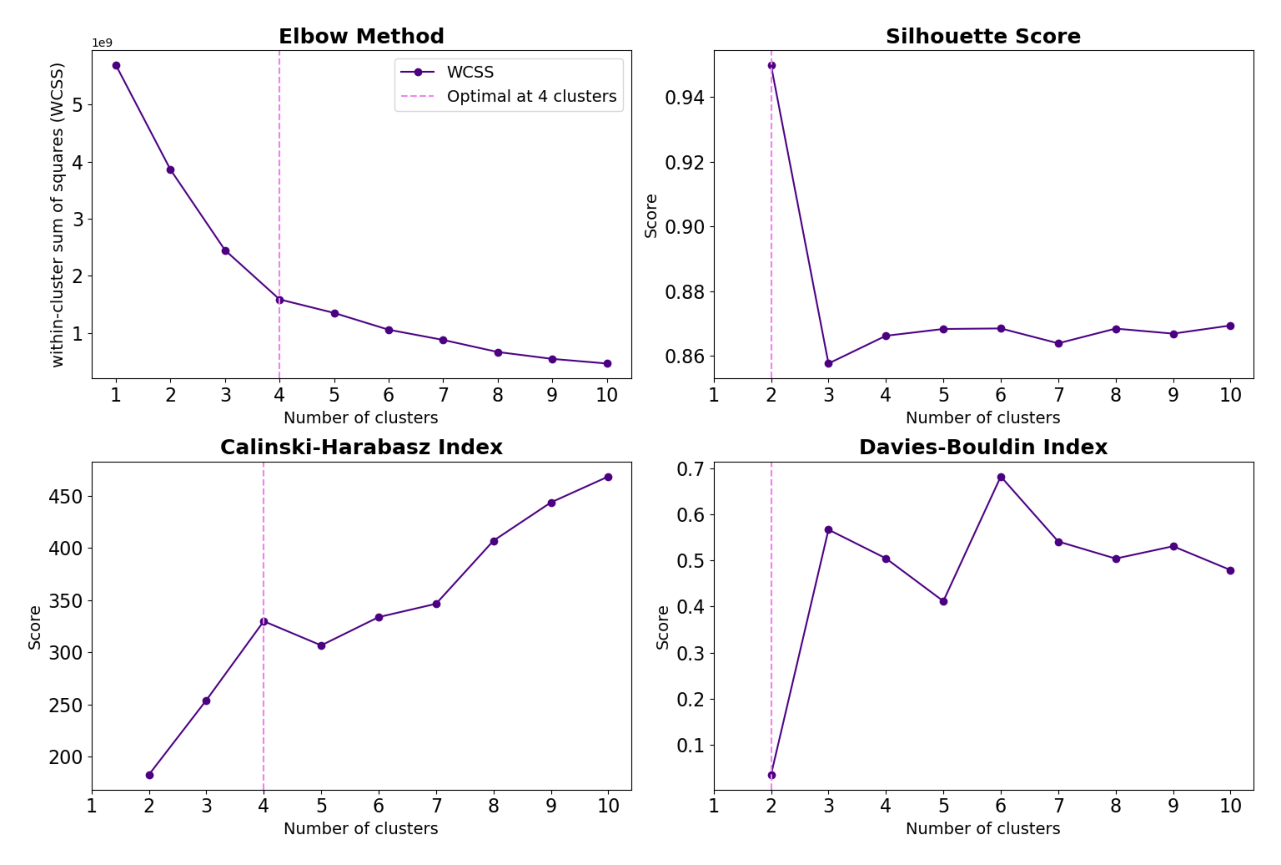}
    \caption{Plot of validation metrics to determine optimal number of
     clusters}
    \label{fig:5.4}
\end{figure}

\begin{figure}
    \centering
    \includegraphics[width=0.75\linewidth]{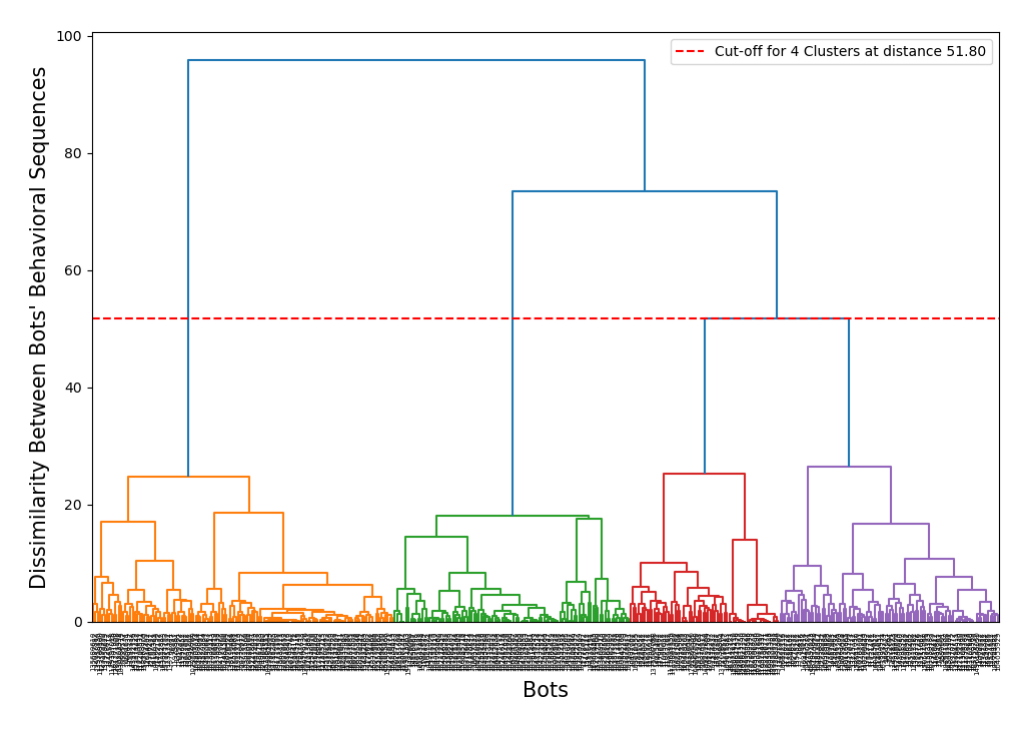}
    \caption{Hierarchical clustering of behaviours into the four family trees}
    \label{fig:5.3}
\end{figure}

Figure~\ref{fig:5.3} shows the dendrogram (the hierarchical tree of bot
families). To determine the number of clusters (families), we examine several
validation metrics (Elbow, Silhouette, Calinski--Harabasz, Davies--Bouldin), as
plotted in Figure~\ref{fig:5.4}/
 Based on these metrics, we select four clusters as a balance
 between parsimony and behavioural granularity.

The resulting cluster set is
\begin{equation}
\textit{Families} = \{C_1, C_2, \dots, C_k\}, \quad k=4,
\label{eq:families}
\end{equation} where each cluster $C_\ell$ is a bot family. 
\subsection{Identification of Bot Families}

The hierarchical tree in Figure~\ref{fig:5.3} shows that the promotional bots
cluster naturally into four families at an appropriate dissimilarity level,
supported by the validation metrics in Figure~\ref{fig:5.4}. While two clusters
also appear reasonable under some metrics (e.g., Silhouette, Davies--Bouldin),
we select four clusters to capture finer behavioural distinctions that are
valuable for understanding and predicting behaviour change.

The four families are:
\begin{itemize}
    \item Family~1: \emph{Unique Tweeters} (129 bots),
    \item Family~2: \emph{Duplicators with URLs} (101 bots),
    \item Family~3: \emph{Content Multipliers} (64 bots),
    \item Family~4: \emph{Informed Contributors} (94 bots).
\end{itemize}

Their feature distributions are summarised visually in Figure~\ref{fig:5.5}.

\subsection{Family-Level Behavioural Characteristics}

\paragraph{Family~1: Unique Tweeters.}

Bots in this family primarily produce original, text-based content. Their
dominant patterns include:
\begin{itemize}
    \item high use of plain tweets (T),
    \item low use of URLs (X), media (M), hashtags (Z $\rightarrow$ ``no
     hashtags''), and emojis (K $\rightarrow$ ``no emojis''),
    \item predominance of unique text (Q) with neutral sentiment (L).
\end{itemize} Overall, this family tends to distribute standalone textual
 updates with minimal engagement features. This profile is visible in the
 feature distribution of Figure~5.5 and in their top blocks in Figure~5.6.

\paragraph{Family~2: Duplicators with URLs.}

This family is characterised by:
\begin{itemize}
    \item systematic use of URLs (U),
    \item frequent text duplication (D),
    \item neutral-to-positive sentiment (L, P),
    \item relatively sparse use of hashtags (often Z) and emojis (K).
\end{itemize} These bots repeatedly push externally linked content, often
 reusing the same text templates. Their block-level patterns include variants
 such as TUMKZDL and TUMKZDP (see Figure~5.6).

\paragraph{Family~3: Content Multipliers.}

Content Multipliers show the richest feature usage:
\begin{itemize}
    \item higher activity across posting actions (T, R, Y) than other families,
     with many retweets and replies,
    \item extensive use of media (I, V), hashtags (H), emojis (J), and URLs(U),
    \item substantial duplication of text (D), with neutral sentiment dominant
     but the highest proportion of negative sentiment (N) among families.
\end{itemize} These bots appear optimised for engagement and amplification,
 using rich media and interaction features intensively. They contribute the
 largest number of unique blocks, especially those involving replies, media,
 emojis, and polarised sentiment (see Figure~5.7).

\paragraph{Family~4: Informed Contributors.}

Bots in this family:
\begin{itemize}
    \item frequently include URLs (U),
    \item mix unique (Q) and duplicated (D) text,
    \item regularly use media (I or V), emojis (J), and hashtags (H),
    \item lean towards positive sentiment (P), with some neutral content (L).
\end{itemize} Their behaviour resembles information-rich promotional accounts
 that aim to be engaging and visually appealing while maintaining a generally
 positive tone.

\subsection{Common and Unique Behavioural Blocks}

Figure~5.6 lists the ten most frequent blocks in each family. Several appear in all families, suggesting a shared core of promotional behaviour:
\begin{itemize}
    \item TXMKZDL: Tweet, no URL, no media, no emoji, no hashtag, duplicated
     text, neutral sentiment;
    \item TUMKZDL: Tweet, URL, no media, no emoji, no hashtag, duplicated text,
     neutral sentiment;
    \item TUMKZEL: Tweet, URL, no media, no emoji, no hashtag, empty text,
     neutral sentiment;
    \item TUMKZQL: Tweet, URL, no media, no emoji, no hashtag, unique text,
     neutral sentiment;
    \item TXMKZQL: Tweet, no URL, no media, no emoji, no hashtag, unique text,
     neutral sentiment;
    \item TUMKZQP: Tweet, URL, no media, no emoji, no hashtag, unique text,
     positive sentiment;
    \item TUMKZDP: Tweet, URL, no media, no emoji, hashtag, duplicated text,
     positive sentiment;
    \item TXMKZDP: Tweet, no URL, no media, no emoji, hashtag, duplicated text,
     positive sentiment.
\end{itemize} These patterns highlight typical promotional structures: a simple
 tweet, often with or without a URL, typically neutral or mildly positive, with
 limited surface decoration (hashtags/emojis) unless reinforcement is desired.

Figure~5.7 focuses on blocks that are unique to a given family. The \emph
{Duplicators with URLs} family has no blocks that are unique to it, suggesting
that its strategy relies more on frequency and proportion of behaviours than on
entirely distinct patterns. In contrast, the \emph{Content Multipliers} family
shows the highest number of unique blocks, predominantly involving:
\begin{itemize}
    \item replies (Y) and retweets (R),
    \item media (I, V),
    \item emojis (J),
    \item duplicated text (D),
    \item negative sentiment (N).
\end{itemize}

\subsection{Characterising Bot Families}

To answer RQ1 and RQ2 we analyse the resulting families from three
perspectives:

\begin{enumerate}[label=(\alph*),leftmargin=*]
    \item \textbf{Family-level feature distributions:} For each family $C_\ell$
     we compute the total count of each encoded letter (feature value) across
     all sequences:
    \begin{equation}
    \text{Count}_L = \sum_{S \in C_\ell} \text{Count}(l, S),
    \label{eq:lettercount}
    \end{equation} then decode letters back to feature values using the inverse
     encoder $E^{-1}$, and use these distributions to qualitatively name and
     describe each family. Figure~5.5 visualises the distribution of
     behavioural features across families.
    
    \item \textbf{Common and unique blocks:} For each family we compute the ten
     most frequent blocks:
    \begin{equation}
    \text{Count}_{\text{block}} = \sum_{S \in C_\ell} \text{Count}(\text
     {block}, S),
    \end{equation} and define
    \begin{equation}
    \text{Top\_Blocks}_{C_\ell} = \text{Sort}_{\text{desc}}\bigl\{\text{Count}_
     {\text{block}}\bigr\}[1\!:\!10].
    \label{eq:top10}
    \end{equation} Figure~5.6 illustrates the ten most common blocks per family,
     and Figure~5.7 highlights blocks that are unique to a single family.
    \begin{figure}
        \centering
        \includegraphics[width=0.75\linewidth]
         {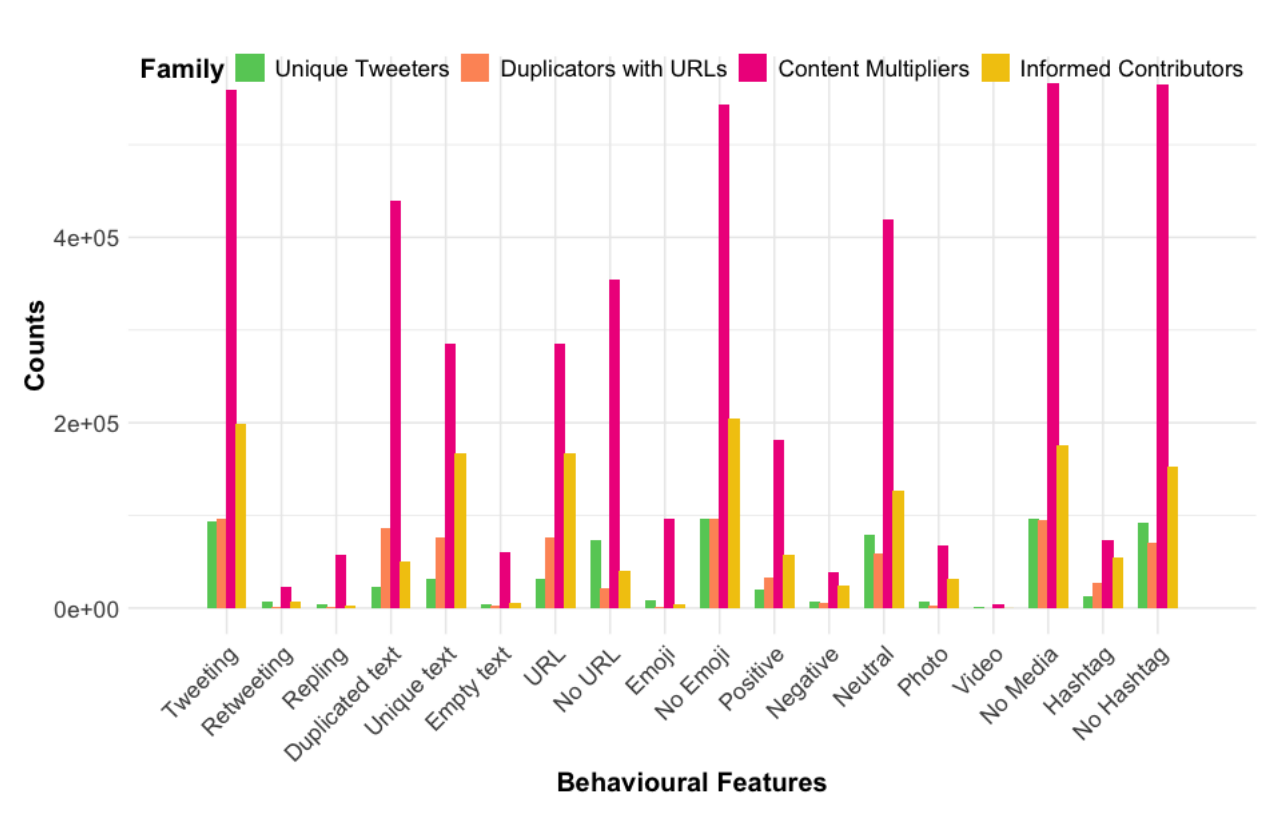}
        \caption{Behavioural and content feature distributions for the four bot
         families}
        \label{fig:5.5}
    \end{figure}
    \begin{figure}
        \centering
        \includegraphics[width=0.75\linewidth]
         {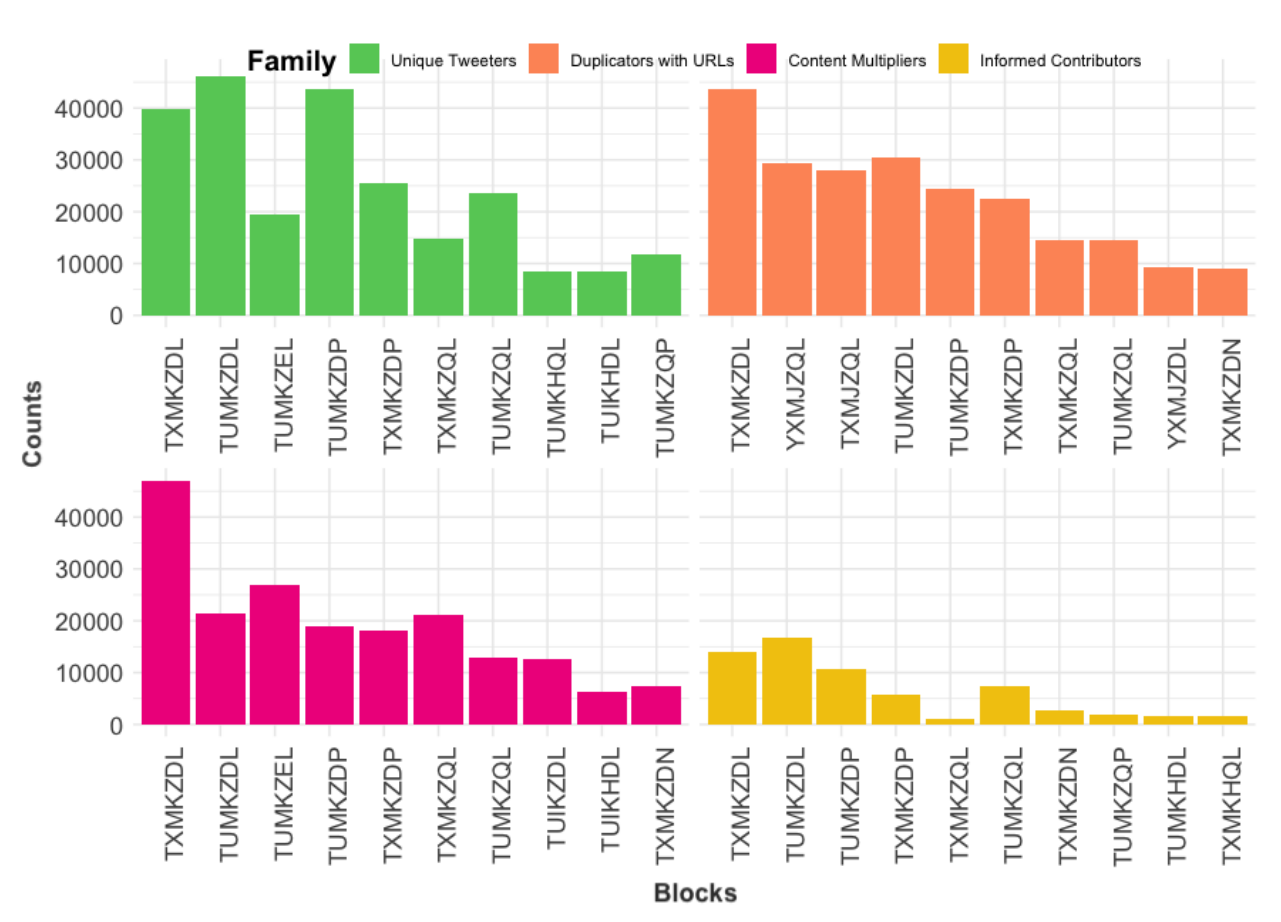}
        \caption{The top ten behavioural blocks for each family}
        \label{fig:5.6}
    \end{figure}
    \begin{figure}
        \centering
        \includegraphics[width=0.75\linewidth]
         {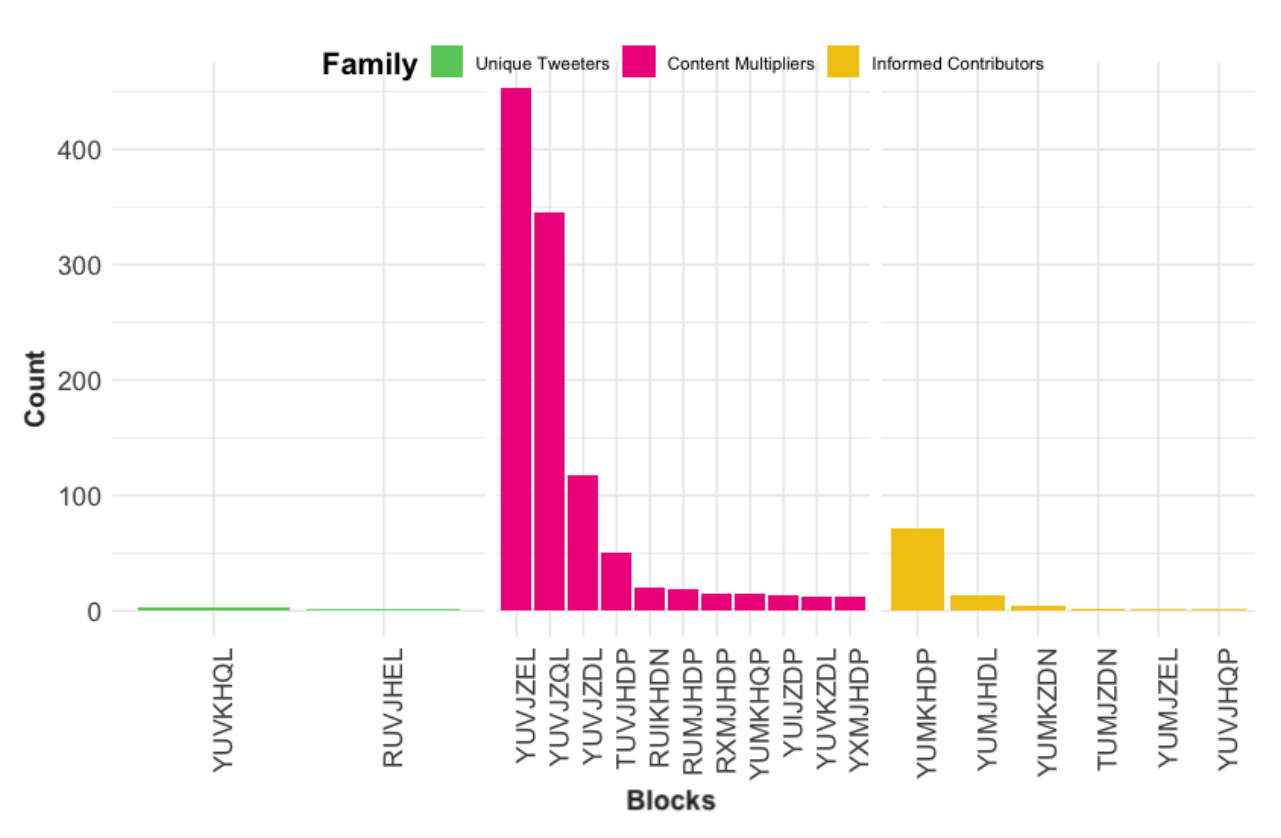}
        \caption{The 10 top unique blocks in each bot family}
        \label{fig:5.7}
    \end{figure}
    \item \textbf{Temporal evolution within families:} To analyse behavioural
     evolution, we segment each bot's sequence into three equal activity
     segments (beginning, middle, final) at the block level. For a sequence
     with $m$ blocks,
    \begin{align}
        \text{Segm}_1 &= \{\text{block}_1, \dots, \text{block}_
         {\lfloor m/3\rfloor}\}, \label{eq:seg1}\\
        \text{Segm}_2 &= \{\text{block}_{\lfloor m/3\rfloor+1}, \dots, \text
         {block}_{2\lfloor m/3\rfloor}\}, \label{eq:seg2}\\
        \text{Segm}_3 &= \{\text{block}_{2\lfloor m/3\rfloor+1}, \dots, \text
         {block}_m\}, \label{eq:seg3}
    \end{align} with any remainder blocks assigned to the third segment.
    
    For each family $C_\ell$, for each of its top five blocks $\text
    {block}_j \in \text{Top\_Blocks}_{C_\ell}$, and for each segment $\text
    {Segm}_k$, we compute the normalised frequency
    \begin{equation}
    \text{Freq}_{\text{block}_j,\text{Segm}_k} =
    \frac{\sum_{S \in C_\ell} \text{Count}(\text{block}_j, \text{Segm}_k)}{\sum_
     {S \in C_\ell} \sum_{\text{block} \in \text{Segm}_k} \text{Count}(\text
     {block}, \text{Segm}_k)}.
    \label{eq:freq}
    \end{equation} Figure~5.8 plots these frequencies, showing the evolution of
     dominant blocks across the three segments for each family.
\end{enumerate}
\begin{figure}
    \centering
    \includegraphics[width=0.75\linewidth]{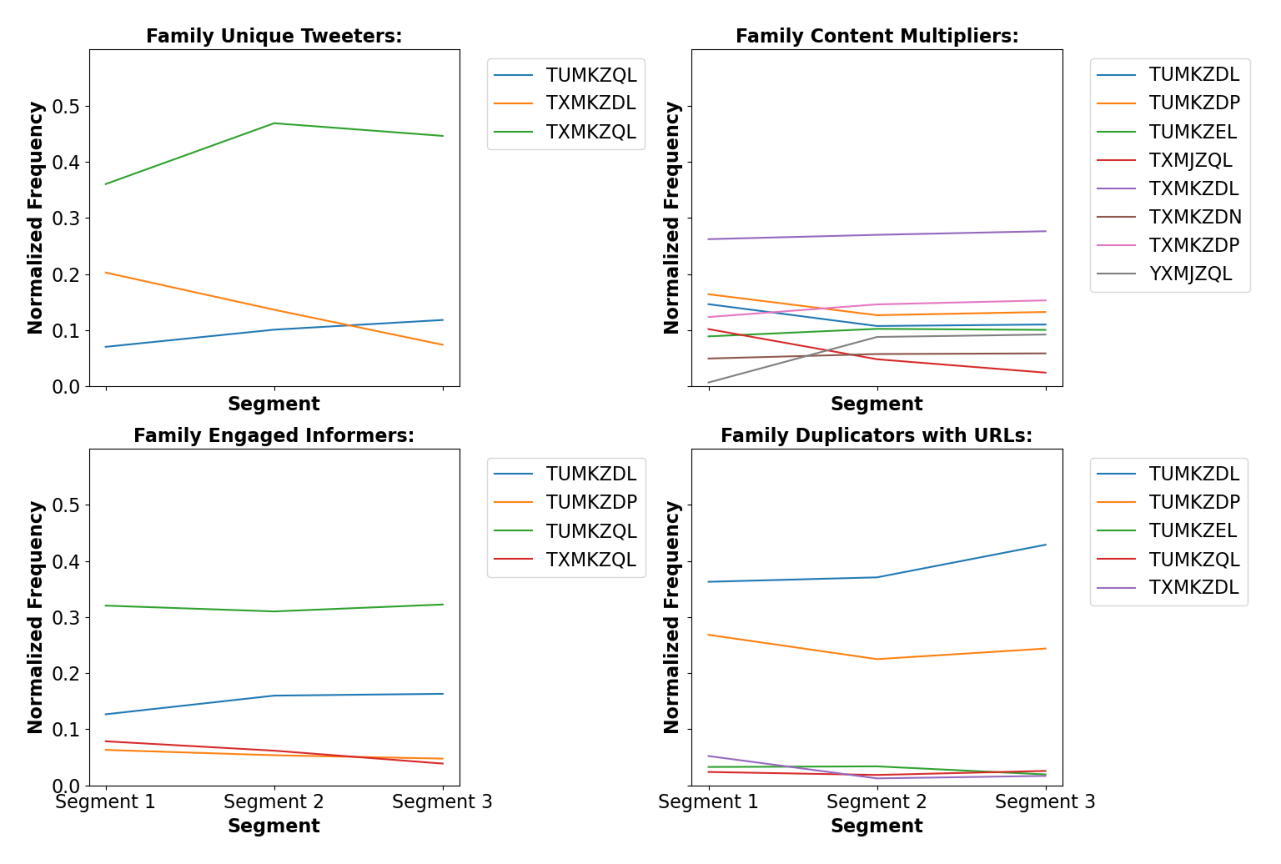}
    \caption{Normalized frequency of the top 5 blocks across three activity
     segments(beginning, middle, and final phases) of bot lifespans for each
     family }
    \label{fig:5.8}
\end{figure}
\subsection{Modelling Behavioural Dynamics via Sequence Alignment and
 Mutations}

To address RQ3, we extend the above framework with a second stage that
explicitly models behavioural change as mutations in the behavioural
sequences.

\subsubsection{Data for Dynamic Analysis}

For each family $C_f$ we consider the set of behavioural sequences
\[ S = \{S_1,S_2,\dots,S_n\}
\]

Aligning sequences within a family allows us to identify \emph
{conserved regions} (behavioural patterns shared across many bots) and \emph
{variable regions} where insertions, deletions or modifications indicate
different strategies. We adapt Multiple Sequence Alignment (MSA), originally
developed for DNA, RNA and protein sequences, to our encoded behavioural
sequences.

Let $S_1,\dots,S_n$ denote the original behavioural sequences of $n$ bots in
family $C_f$. The MSA procedure produces aligned sequences
\begin{equation}
\hat{S} = \{\hat{S}_1,\hat{S}_2,\dots,\hat{S}_n\},
\end{equation} where each $\hat{S}_i$ is derived from $S_i$ by introducing gap
 symbols ``-'' so that all aligned sequences have the same length. We use the
 MAFFT tool to perform MSA because of its scalability and applicability to
 long, non-biological sequences.

Directly aligning sequences of very different lengths can bias the alignment
towards longer sequences and over-pad shorter ones. To mitigate this, we first
cluster sequences by length within each family using quartiles of the length
distribution.

For each family $C_f$:
\begin{enumerate}[leftmargin=*]
    \item Compute sequence lengths $L_i = |S_i|$ and empirical quartiles $Q_1,
     Q_2, Q_3$.
    \item Partition sequences into four clusters
    \[
        \text{group}_1 = \{S_i : L_i < Q_1\},\;
        \text{group}_2 = \{S_i : Q_1 \le L_i < Q_2\},
    \]
    \[
        \text{group}_3 = \{S_i : Q_2 \le L_i < Q_3\},\;
        \text{group}_4 = \{S_i : L_i \ge Q_3\}.
    \]
    \item Within each cluster $\text{group}_k$, pad all sequences to the length
     of the longest sequence in that cluster. Let
    \[ L^{(k)}_{\max} = \max\{|S_i| : S_i \in \text{group}_k\}.
    \] The padded sequence $\hat{S}_i$ is
    \begin{equation}
    \hat{S}_i =
    \begin{cases} S_i, & |S_i| = L^{(k)}_{\max},\\[3pt] S_i \;\text
     {followed by } L^{(k)}_{\max} - |S_i| \text{ gap symbols}, & |S_i| < L^{
     (k)}_{\max},
    \end{cases}
    \end{equation} and the padded cluster is
    \begin{equation}
    \widehat{\text{group}}_k = \{\hat{S}_i : S_i \in \text{group}_k\}.
    \end{equation}
    \item Apply MSA separately to each padded cluster:
    \[
    \hat{S}^{(k)} = \text{MSA}(\widehat{\text{group}}_k).
    \]
    \item Merge the aligned clusters to form the final aligned family:
    \begin{equation}
        \hat{C}_f = \bigcup_{k=1}^{4} \hat{S}^{(k)}.
    \end{equation}
\end{enumerate}

MSA aligns all sequences in a family simultaneously, reveals conserved and
variable regions, and exposes mutation hotspots that would be difficult to
detect from pairwise comparisons alone.

\subsubsection{Detecting Behavioural Mutations}

\begin{figure}
    \centering
    \includegraphics[width=1\linewidth]{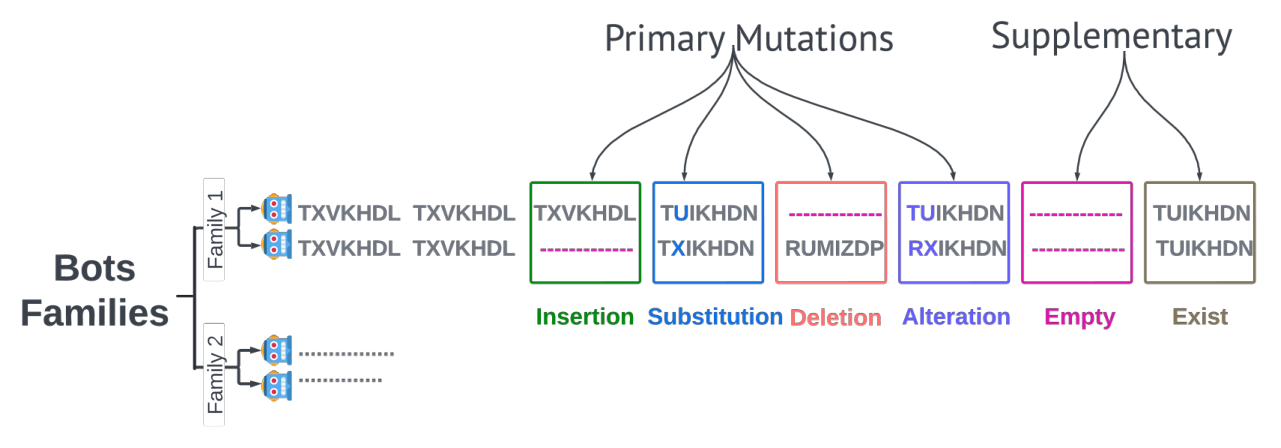}
    \caption{Alignment process for behavioural sequences within each bot
     family}
    \label{fig:6.3}
\end{figure}

Given aligned sequences $\hat{S} = \{\hat{S}_1,\dots,\hat{S}_n\}$ for a family
$\hat{C}_f$, we detect behavioural changes by comparing aligned blocks position
by position across pairs of sequences (Figure~\ref{fig:6.3}). Recall that each
original block is a 7-letter motif (Equation~\ref{eq:block}); alignment may
introduce gaps into these blocks, but we still treat each aligned 7-letter
group (including possible gaps) as a block.

For every pair of aligned sequences $\hat{S}_i,\hat{S}_j$ in $\hat{C}_f$ and
each aligned block position $k$, we consider the aligned blocks $\text{block}^{
(i)}_k$ and $\text{block}^{(j)}_k$. If the blocks differ, we classify the
difference into one of several mutation types. Formally, the set of detected
mutations for family $\hat{C}_f$ is
\begin{equation}
\Delta_f =
\bigcup_{i,j \in \{1,\dots,n\},\, i\neq j}
\{(k, \delta_k) \mid \delta_k = \text{MutationType}(\text{block}^{(i)}_k, \text
 {block}^{(j)}_k),\ \text{block}^{(i)}_k \neq \text{block}^{(j)}_k\},
\label{eq:mutset}
\end{equation} where $\delta_k$ is the mutation type at position $k$ and $\text
 {MutationType}(\cdot,\cdot)$ implements the classification logic summarised
 below. We distinguish the following mutation types:
\begin{itemize}
    \item \textbf{Empty:} both aligned blocks contain only gaps, so no behaviour
     is present to compare.
    \item \textbf{Insertion:} a new block appears in one sequence but not in the
     other at position $k$, and this block has not been seen previously in that
     sequence. This corresponds to a new behavioural pattern being introduced
     (block vs. gaps).
    \item \textbf{Deletion:} a block present earlier in both sequences is now
     absent in one sequence (gaps vs. a previously seen block), indicating that
     a previously used behaviour has been discontinued.
    \item \textbf{Substitution:} the two blocks are identical except for a
     single letter, so exactly one behavioural feature (e.g., URL presence) has
     changed while the rest of the structure is preserved.
    \item \textbf{Alteration:} multiple letters differ between the two blocks,
     but at least half of the letters still match (similarity $\geq 50\%$) and
     the block has been observed previously in one of the sequences. This
     reflects a partial modification of an existing pattern.
    \item \textbf{Identity:} a block appears at different positions in the two
     sequences, indicating that an existing behavioural pattern has been
     re-used but in a different context or order.
\end{itemize}

\begin{figure}
    \centering
    \includegraphics[width=1\linewidth]{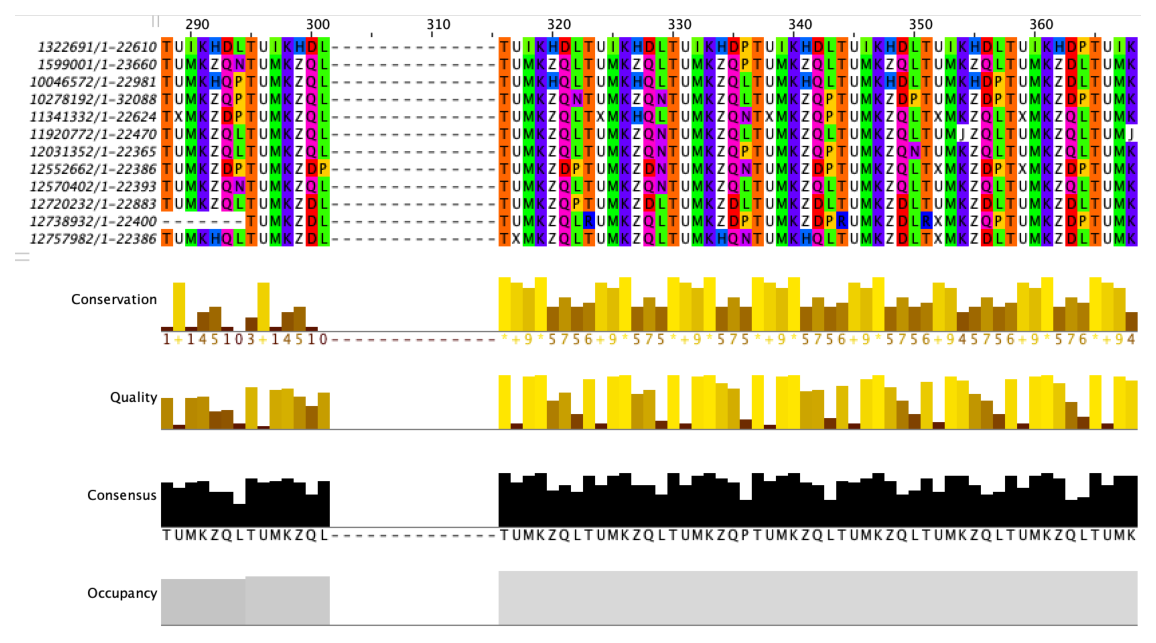}
    \caption{Graphical representation of aligned bot sequences}
    \label{fig:6.5}
\end{figure} Figure~\ref{fig:6.5} shows an example of aligned sequences
 visualised using the Jalview tool: each row is a bot sequence, coloured by
 letter; the consensus and conservation scores below the alignment highlight
 the most common behaviour at each position.

The classification distinguishes between genuinely new blocks
(insertions), partial modifications of previously known blocks
(alterations), single-feature changes (substitutions), and simple reuse
(exist). The 50\% threshold separating alterations from insertions follows
practices in DNA fingerprinting and related sequence-comparison work, where
similarity thresholds are used to distinguish self-like patterns from more
distant ones; it provides a simple and interpretable cut-off that can be
refined in future work.

\subsubsection{Mutation Statistics and Visualisation}

\begin{figure}[htb]
\centering
\includegraphics[width=1\linewidth]{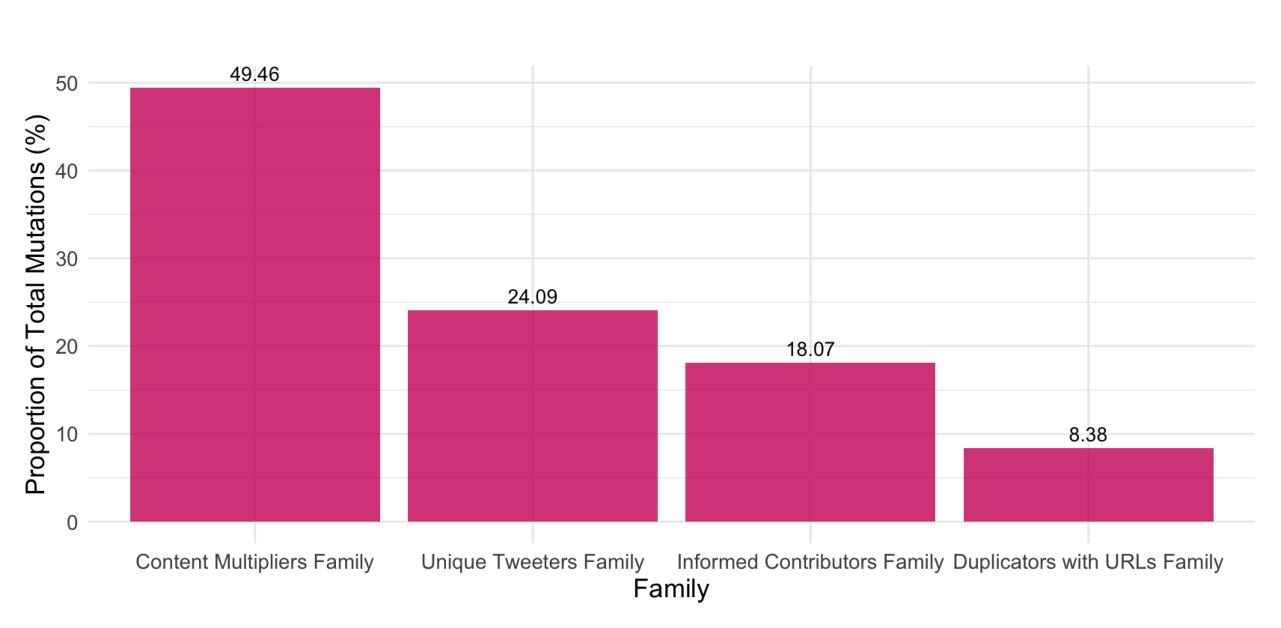}
\caption{Mutation frequencies across families}
\label{fig:6.6}
\end{figure}

To interpret the detected mutations, we compute several summary statistics
within each family $\hat{C}_f$:

\paragraph{Mutation-type frequencies.} For each mutation type $\delta$
 (in particular, insertions, deletions, substitutions, alterations, and the
 supplementary category ``identity'' (match/no-change)) we compute its proportion:
\begin{equation} P(\delta \mid \hat{C}_f) =
\frac{\text{Count}(\delta, \hat{C}_f)}{\text{TotalMutations}(\hat{C}_f)},
\label{eq:mutfreq}
\end{equation} where $\text{Count}(\delta,\hat{C}_f)$ is the number of times
 mutation type $\delta$ occurs in family $\hat{C}_f$, and $\text
 {TotalMutations}(\hat{C}_f)$ is the total number of detected mutations in that
 family. Figure \ref{fig:6.6} plots these frequencies across families.

\begin{figure}[htb]
\centering
\includegraphics[width=1\linewidth]{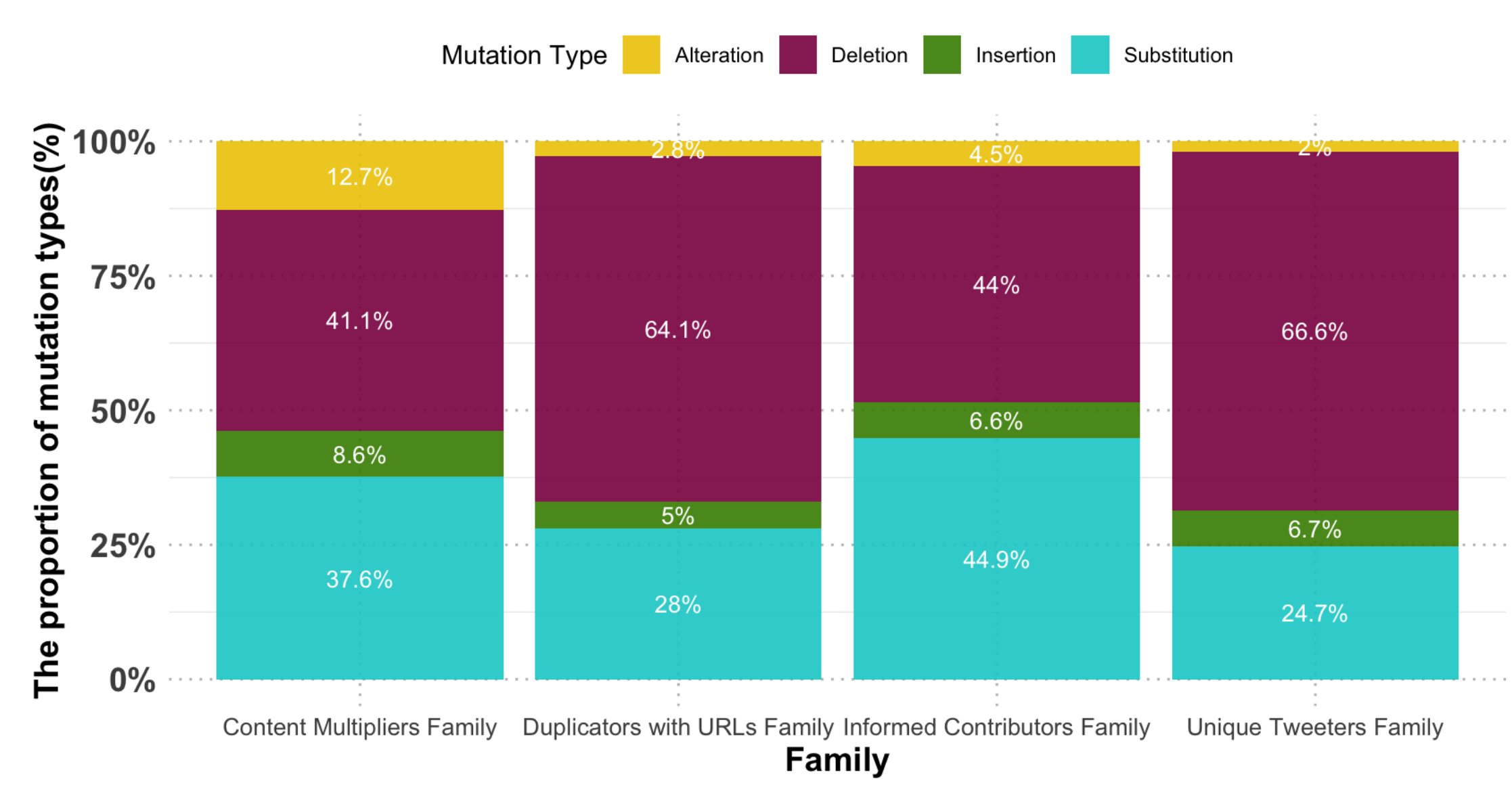}
\caption{The distribution of the primary mutation types across bot families}
\label{fig:6.7}
\end{figure}
\begin{figure}
    \centering
    \includegraphics[width=1\linewidth]{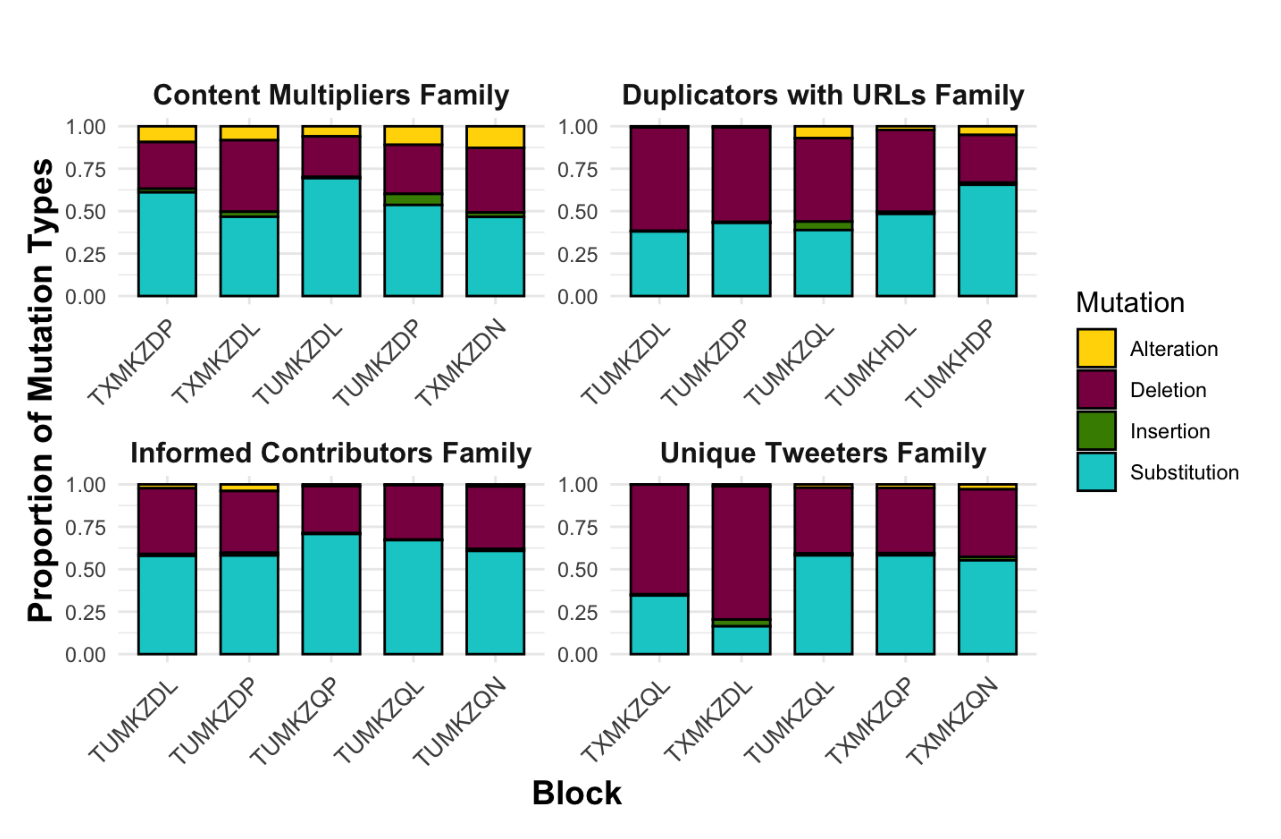}
    \caption{Proportion of mutation types for the top 5 most mutated blocks in
     each bot family}
    \label{fig:6.9}
\end{figure}
\begin{figure}
    \centering
    \includegraphics[width=1\linewidth]{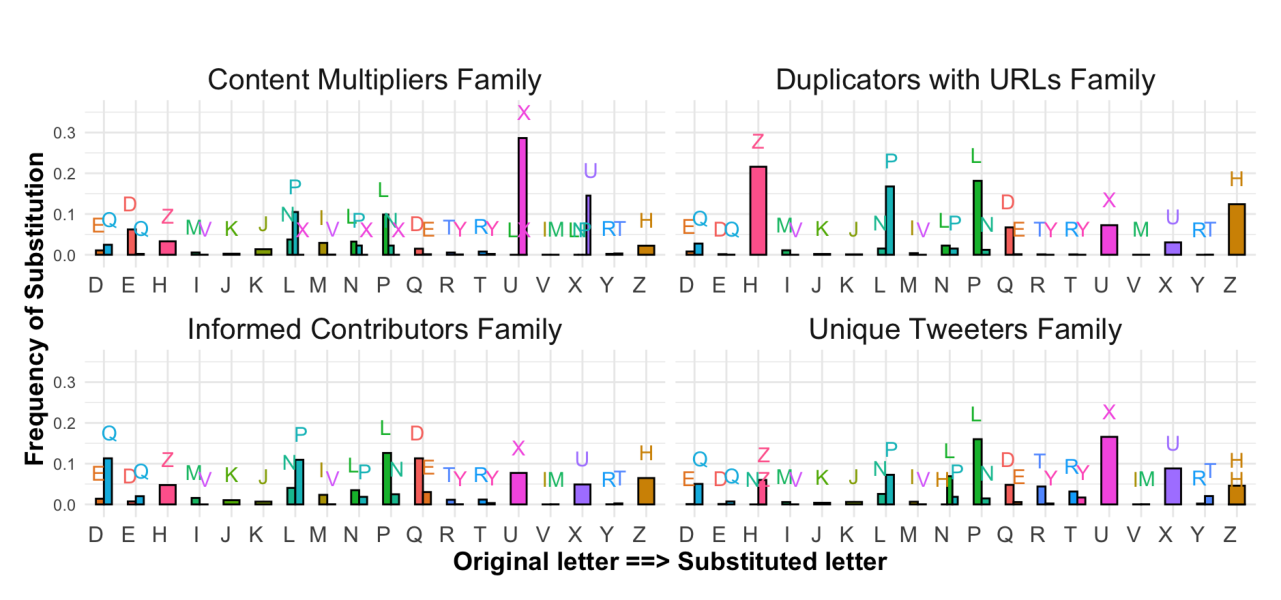}
    \caption{The frequency of substituted letters from letter 1 of block 1 to
     block 2 within each bot’s family.}

    \label{fig:6.10}
\end{figure}
\begin{figure}
    \centering
    \includegraphics[width=1\linewidth]{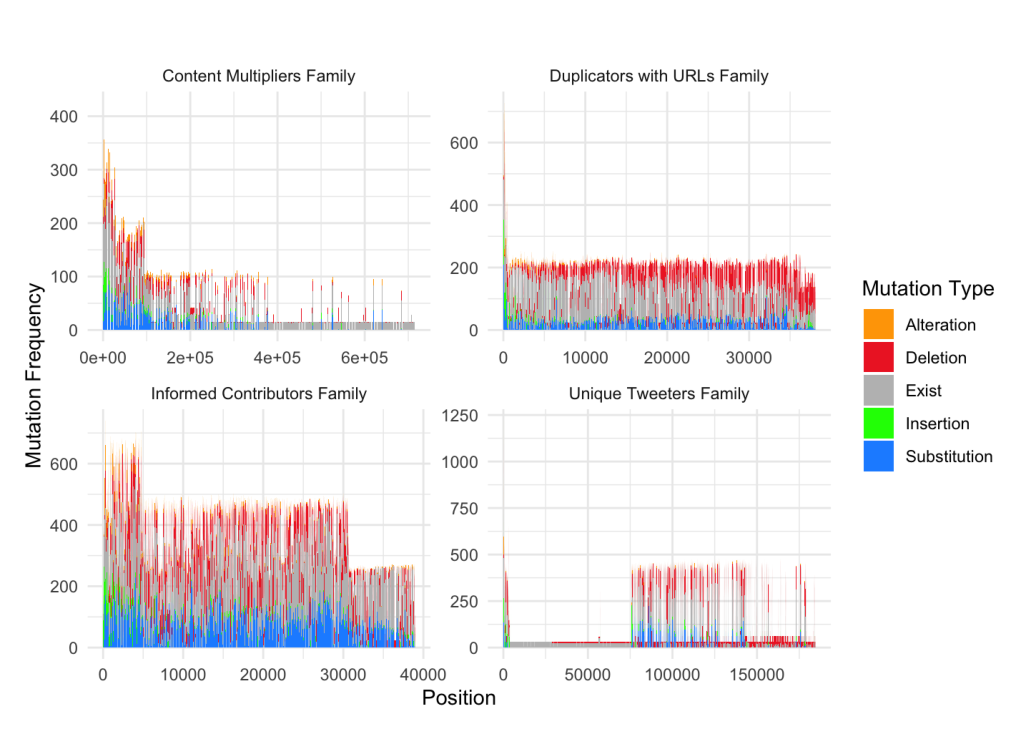}
    \caption{Mutation hotspot analysis across bot families}
    \label{fig:6.11}
\end{figure}

\paragraph{Blocks most prone to mutation.} For each block pattern $\text
 {block}_k$ in $\hat{C}_f$, we compute the total number of mutations affecting
 it and the proportion of each mutation type:
\begin{equation}
\text{Proportion}(\delta,\text{block}_k,\hat{C}_f) =
\frac{\text{Count}(\delta,\text{block}_k,\hat{C}_f)}{\text{TotalMutations}(\text
 {block}_k,\hat{C}_f)},
\label{eq:blockfreq}
\end{equation} where $\text{TotalMutations}(\text{block}_k,\hat{C}_f)$ sums all
 mutation types affecting $\text{block}_k$. Ranking blocks by $\text
 {TotalMutations}(\text{block}_k,\hat{C}_f)$ yields the blocks most prone to
 mutation; Figure~\ref{fig:6.9} shows the top five per family.
\paragraph{Substitution patterns between feature values.} For substitution
 mutations, where a single letter $l_1$ is replaced by $l_2$
 (e.g., U$\rightarrow$X, indicating a move from URL to no-URL), we compute
\begin{equation}
\text{Frequency}(l_1 \to l_2,\hat{C}_f) =
\frac{\text{Count}(l_1 \to l_2,\hat{C}_f)}{\text{TotalBlocks}(\hat{C}_f)},
\label{eq:subfreq}
\end{equation} where $\text{TotalBlocks}(\hat{C}_f)$ is the total number of
 aligned blocks in the family. Figure~\ref{fig:6.10} reports these substitution
 frequencies for each family.
\paragraph{Mutation hotspots.} Finally, we quantify how heavily each aligned
 position $k$ is affected by mutations via a hotspot score:
\begin{equation}
\text{Hotspot\_Score}(k,\hat{C}_f) =
\frac{\sum_{\delta} \text{Count}(\delta, k,\hat{C}_f)}{\text{TotalMutations}
 (\hat{C}_f)},
\label{eq:hotspot}
\end{equation} where $\text{Count}(\delta,k,\hat{C}_f)$ counts mutations of type
 $\delta$ at position $k$. Positions with high hotspot scores are locations
 where behaviour changes frequently; Figure~\ref{fig:6.11} shows hotspot
 profiles for each family.
 Visual inspection of aligned sequences and their consensus, using
 Jalview (Figure~\ref{fig:6.5}), complements these quantitative measures by
 revealing where gaps, conserved regions and variable regions occur along the
 sequences.

\section{Bot Family Results}

\subsection{Behavioural Evolution Within Families}

Figure~5.8 visualises the normalised frequency of the top five blocks per family
across three segments of each bot's lifespan (beginning, middle, final). The
trajectories of these lines illustrate stable, increasing, and decreasing
behavioural trends.

\paragraph{Stable trends.}

Some blocks remain consistently frequent across all segments, indicating stable
behavioural strategies. For example, in the \emph{Unique Tweeters} family,
TXMKZQL (tweet, no URL, no media, no emoji, no hashtag, unique text, neutral
sentiment) remains high and relatively flat over time, suggesting persistent
emphasis on original text updates.

\paragraph{Decreasing trends.}

Certain patterns decline as bots age. In the \emph{Unique Tweeters} family,
TXMKZDL (duplicated text without URL or media) decreases from the first to the
third segment, indicating a gradual shift away from pure duplication.

\paragraph{Increasing trends.}

Other blocks show sustained growth. In the \emph{Duplicators with URLs} family,
TUMKZDL (duplicated text with URL) steadily increases across all segments,
signalling an intensifying emphasis on duplicative URL-based promotion. In
the \emph{Informed Contributors} family, blocks with URLs and richer
embellishments show modest upward trends.

\paragraph{Family-specific evolution.}

The evolution of a given block can differ markedly between families. For
example, TUMKZDL behaves differently in two families:
\begin{itemize}
    \item In \emph{Content Multipliers}, its frequency decreases early and then
     stabilises, suggesting a shift towards more complex engagement patterns
     rather than simple duplicated URL tweets.
    \item In \emph{Duplicators with URLs}, its frequency increases steadily
     across all segments, reinforcing their role as heavy URL-based
     duplicators.
\end{itemize} These family-specific trajectories help answer RQ2.B by showing
 that members of a given family tend to evolve in similar ways, but these
 trajectories differ across families.

\subsection{Global Mutation Frequencies Across Families}

The MSA-based analysis extends these results by quantifying how often different
types of behavioural mutations occur within each family (Figure~\ref
{fig:6.6}). The total number of mutations is not uniform across families:
\begin{itemize}
    \item The \emph{Content Multipliers} family contributes almost half of all
     detected mutations ($\approx 49.46\%$).
    \item \emph{Unique Tweeters} account for around $24\%$.
    \item \emph{Informed Contributors} for about $18\%$.
    \item \emph{Duplicators with URLs} for the remaining $8\%$.
\end{itemize}

Across families, primary mutation types follow a consistent ranking (Figure~6.7):
\begin{itemize}
    \item \textbf{Deletions} are the most common mutation type overall, with an
     average proportion of $53.96\%$. They are particularly dominant for \emph
     {Unique Tweeters} (66.58\%) and \emph{Duplicators with URLs} (64.14\%).
    \item \textbf{Substitutions} are the second most frequent type, averaging
     $33.81\%$. They are most prominent for \emph{Informed Contributors}
     (44.91\%) and least for \emph{Unique Tweeters} (24.69\%).
    \item \textbf{Insertions} are rare, with an average of $6.73\%$, and only
     modestly higher in the \emph{Informed Contributors} family (8.74\%).
    \item \textbf{Alterations} are the least common primary mutation, averaging
     $5.50\%$. They are noticeably more frequent in \emph{Content Multipliers}
     (12.68\%) than in the other families (all under $4.5\%$).
\end{itemize}

Figure~6.8 compares the supplementary category ``Identity'' with the primary
mutation types. In all families, ``Identity'' accounts for a very large share of
events, often around half of all counted cases. The \emph{Content Multipliers}
family has the highest proportion of ``Identity'' events (57.5\%), reflecting
highly recurrent patterns that reappear in multiple positions over the aligned
sequences.

\subsection{Blocks Most Prone to Mutation}

Figure~6.9 shows the top five blocks most affected by mutations in each family.
These blocks are also prominent in the static top-10 lists of Figure~5.6,
showing that core patterns are not only frequent but also heavily modified over
time.

Some blocks, such as TUMKZDL, TUMKZDP and TUMKZQL, repeatedly appear among the
most mutated patterns across multiple families. They are primarily affected by
substitutions and deletions; insertions and alterations occur but at lower
rates. The \emph{Content Multipliers} family stands out with slightly higher
proportions of insertions and alterations on these blocks than other families,
consistent with its more complex and adaptive behaviour.

\subsection{Most Altered Features in Substitution Mutations}

Figure~6.10 breaks down substitution mutations by letter pairs within each
family, highlighting which features are most often changed.

\paragraph{Content Multipliers.}

In the \emph{Content Multipliers} family, substitutions concentrate on:
\begin{itemize}
    \item URL presence: U$\to$X (URL $\to$ no URL) and X$\to$U (no URL $\to$
     URL),
    \item sentiment: L$\to$P (neutral $\to$ positive) and P$\to$L(positive $\to$
     neutral),
    \item text status: E$\to$D (empty $\to$ duplicated),
    \item hashtags: Z$\to$H (no hashtags $\to$ hashtags),
    \item sentiment dampening: N$\to$L (negative $\to$ neutral),
    \item media: M$\to$I (no media $\to$ image) and M$\to$V (no media $\to$
     video),
    \item emojis: K$\to$J (no emoji $\to$ emoji).
\end{itemize} This profile reflects frequent toggling of URLs, sentiment, and
 surface features such as media, hashtags and emojis to maximise engagement.

\paragraph{Duplicators with URLs.}

In the \emph{Duplicators with URLs} family, substitutions emphasise:
\begin{itemize}
    \item hashtags: H$\to$Z (hashtags $\to$ no hashtags) and Z$\to$H(no hashtags
     $\to$ hashtags),
    \item sentiment: P$\to$L (positive $\to$ neutral) and L$\to$P (neutral $\to$
     positive),
    \item URL presence: U$\to$X (URL $\to$ no URL) and X$\to$U (no URL $\to$
     URL),
    \item text duplication: Q$\to$D (unique $\to$ duplicated text).
\end{itemize} These substitutions suggest that this family primarily fine-tunes
 visibility via hashtags and tone via sentiment, while occasionally adjusting
 URL usage and duplication.

\paragraph{Informed Contributors.}

For the \emph{Informed Contributors} family, the most frequent substitutions
are:
\begin{itemize}
    \item sentiment: P$\to$L (positive $\to$ neutral) and L$\to$P (neutral $\to$
     positive),
    \item duplication: D$\to$Q and Q$\to$D (duplicated $\leftrightarrow$ unique
     text),
    \item URLs: U$\to$X and X$\to$U,
    \item hashtags: Z$\to$H and H$\to$Z,
    \item sentiment polarisation: L$\to$N (neutral $\to$ negative) and N$\to$L
     (negative $\to$ neutral).
\end{itemize} This mixture emphasises careful modulation of sentiment,
 duplication and engagement features, consistent with an ``informed'' style
 seeking credibility and reach.

\paragraph{Unique Tweeters.}

In the \emph{Unique Tweeters} family, substitutions focus on:
\begin{itemize}
    \item URLs: U$\to$X and X$\to$U,
    \item sentiment: P$\to$L, L$\to$P and N$\to$L,
    \item hashtags: H$\to$Z and Z$\to$H,
    \item duplication: D$\to$Q and Q$\to$D,
    \item posting action: R$\to$T (retweet $\to$ tweet).
\end{itemize} These substitutions show that even this comparatively simple
 family adjusts URLs, sentiment and hashtags over time while maintaining a
 predominantly original tweeting style.

\subsection{Mutation Hotspots}

Figure~6.11 presents mutation hotspot profiles across the four families. The
distribution of hotspot scores reveals distinct patterns:

\begin{itemize}
    \item \textbf{Content Multipliers} show widespread mutation activity, with
     pronounced peaks near the beginning of sequences and non-trivial activity
     throughout later positions. This reflects their continuous and adaptive
     engagement strategy, involving frequent modifications to URLs, sentiment,
     media and other features.
    \item \textbf{Duplicators with URLs} exhibit concentrated hotspots near the
     start of their sequences, corresponding to early blocks where URLs and
     sentiments are frequently tweaked. Later positions are more stable,
     consistent with repetitive use of established patterns.
    \item \textbf{Informed Contributors} have relatively high mutation
     frequencies at early positions, then more evenly distributed activity
     thereafter, indicating balanced adaptation of foundational features
     (e.g., sentiment, hashtags) and ongoing adjustments later in the
     sequence.
    \item \textbf{Unique Tweeters} display sparse mutation hotspots, with modest
     peaks at the beginning and isolated spikes later on. This is consistent
     with a minimalist, originality-focused strategy that makes occasional
     adjustments rather than frequent changes.
\end{itemize}

\section{Bot Families Discussion}

\subsection{Sequence-Based Encoding and Family Structure}

The results provide empirical support for representing promotional bot behaviour
as unified sequences of multi-feature blocks. Compared with previous work that
either:
\begin{itemize}
    \item encodes only a single feature into a sequence (e.g., posting actions),
     or
    \item maintains multiple independent sequences for different features,
\end{itemize} the unified block representation:
\begin{enumerate}[label=(\roman*),leftmargin=*]
    \item preserves the joint configuration of features at the post level;
    \item yields a single, interpretable sequence per bot;
    \item supports both granular, post-level interpretation and global,
     sequence-wide analysis via $k$-mers.
\end{enumerate}

Hierarchical clustering over cosine-based frequency vectors produces a coherent
family structure (Figure~5.3), which is further validated by multiple cluster
quality metrics (Figure~5.4). Four families provide a good compromise between
simplicity and the ability to capture distinct behavioural strategies, directly
addressing RQ2.A.

\subsection{Similarity and Uniqueness of Behavioural Patterns}

The presence of shared core blocks across all families (Figure~5.6) suggests
that promotional bots, regardless of specific strategy, rely on a small set of
characteristic patterns: tweeting either unique or duplicated text, with or
without URLs, usually in neutral or mildly positive tone, and often without
heavy use of hashtags or emojis. These blocks may serve as strong candidates
for core promotional-bot signatures in detection systems.

At the same time, family-specific unique blocks (Figure~5.7) highlight the
discriminative power of features such as:
\begin{itemize}
    \item retweeting and replying behaviour,
    \item media usage,
    \item emojis,
    \item polarised sentiment.
\end{itemize} For example, \emph{Content Multipliers} are distinguished by
 reply/retweet-heavy patterns with media and polarised sentiment, whereas \emph
 {Informed Contributors} show unique reply patterns without media but with
 clear sentiment. The absence of unique blocks in \emph{Unique Tweeters}
 and \emph{Duplicators with URLs} indicates that their distinctiveness is more
 about the \emph{relative} use of shared patterns than about exclusive
 behaviours.

\subsection{Within-Family Behavioural Evolution}

Coarse life-cycle segmentation (Figure~5.8) shows that strategies evolve over time. Importantly, the
evolution is coherent within families:
\begin{itemize}
    \item \emph{Unique Tweeters} gradually shift from duplicated to unique text,
     with some growth in URL usage, consistent with a move from template-like
     promotion towards more original content.
    \item \emph{Duplicators with URLs} increase reliance on duplicated URL
     content, becoming more aggressively promotional over time.
    \item \emph{Content Multipliers} maintain high use of engagement features
     and adjust relative usage of simple URL-duplication blocks, favouring more
     complex interaction patterns.
    \item \emph{Informed Contributors} evolve in parallel across duplicated and
     unique content, and across posts with and without URLs, reflecting a
     balanced but increasingly rich promotional strategy.
\end{itemize}

These patterns support RQ2.B: bots within the same family follow similar
evolutionary trajectories, while different families exhibit distinct
trajectories. This has implications for detection: rather than treating bots as
static points in feature space, it may be more effective to model trajectories
within a family-specific behavioural manifold.

\subsection{Dynamic Mutation Patterns Across Families}

Our mutation framework refines this picture by tracking how specific patterns
change and where changes occur.

First, the dominance of deletions and substitutions (Figure~\ref
{fig:6.7} suggests that bots most often adapt by modifying or dropping existing
patterns rather than by introducing entirely new ones. Insertions are rare;
genuine novelty in behavioural blocks is infrequent. Alterations
(multi-feature modifications) are also uncommon except in the most complex
family, \emph{Content Multipliers}, where more intricate changes occur.

Second, mutation-type profiles align with each family’s static behavioural signature:
\begin{itemize}
    \item \emph{Unique Tweeters} rely heavily on deletions and simple feature
     substitutions, consistent with a strategy of simplifying content and
     making lightweight adjustments to URLs, sentiment and hashtags.
    \item \emph{Duplicators with URLs} mainly adjust sentiment and hashtags
     while preserving their core duplicated-URL structure, which matches their
     role as repetitive link promoters with slight variations.
    \item \emph{Content Multipliers} exhibit the richest mutation mix, including
     more alterations and a high proportion of ``identity'' events, reflecting
     sustained re-use and re-configuration of complex patterns.
    \item \emph{Informed Contributors} balance deletions and substitutions,
     especially of sentiment, duplication status and engagement features,
     reflecting nuanced adaptation of a mixed content strategy.
\end{itemize}

The substitution statistics in Figure~\ref{fig:6.10} reinforce this
interpretation. Across families, the most common substitutions involve:
\begin{itemize}
    \item toggling URLs (U$\leftrightarrow$X),
    \item modulating sentiment (L$\leftrightarrow$P and L$\leftrightarrow$N),
    \item switching between unique and duplicated text (Q$\leftrightarrow$D),
    \item turning hashtags and emojis on or off (Z$\leftrightarrow$H, K$\to$J),
    \item occasionally switching media types (M$\to$I, M$\to$V).
\end{itemize} These are precisely the features that can be modified without
 changing the core topical content, and thus are natural levers for bots
 seeking to adapt while remaining on-message.

Mutation hotspots (Figure~\ref{fig:6.11}) provide a complementary view. Families
with complex, high-activity strategies (\emph{Content Multipliers}, \emph
{Informed Contributors}) exhibit broader and more distributed hotspots,
indicating adaptation spread throughout their behavioural timeline. In
contrast, families with simpler strategies (\emph{Unique Tweeters}, \emph
{Duplicators with URLs}) show more localised hotspots, often concentrated at
the beginning of sequences, reflecting early experimentation followed by
stabilisation.

\subsection{Stylised empirical regularities}

Across the analyses, several ``stylised facts'' emerge that generalise beyond this dataset. First, behavioural families combine a shared \emph{core} of persistent blocks with a family-specific periphery that differentiates engagement strategies. Second, change is strongly \emph{asymmetric}: deletions and substitutions dominate, while insertions are rare, suggesting adaptation often toggles or re-weights existing components (e.g., URLs, duplication, sentiment, hashtags) rather than introducing new activity types. Third, change is \emph{localised in time}: some families concentrate mutation hotspots early (rapid initial adaptation), while others spread change across their life-cycle. Finally, external triggers (e.g., seasonal events) act as family-specific perturbations, producing distinct and partially repeatable responses in both activity and mutation patterns.

Taken together, these regularities motivate analysis and monitoring pipelines
that treat automated accounts not as independent points but as members of
evolving behavioural lineages, and they provide interpretable targets for drift
monitoring and family-conditioned prediction in adversarial settings.

\subsection{Comparative Positioning Against Related Work}

Table~\ref{tab:comparison-methodologies} situates our approach against three strands of related work: DNA-based entropy methods, machine-learning classifiers, and NLP-heavy fake-news/bot pipelines. Most prior work targets detection and treats behaviour as static feature vectors, or summarises sequences with global statistics (e.g., entropy). In contrast, we model within-account change explicitly via alignment and mutation operators, which localises behavioural edits over time and supports family-level comparisons.

\begin{table}[htb!]
\centering
\caption{Comparison of Methodologies for Bot Behaviour Analysis}
\label{tab:comparison-methodologies}
\begin{tabular}{|p{0.16\textwidth}|p{0.2\textwidth}|p{0.2\textwidth}|p
 {0.2\textwidth}|p{0.2\textwidth}|}
\hline
\textbf{Dimension} & \textbf{Our Work} & \textbf{Gilmary~[65]} & \textbf
 {Cheng et al.~[27]} & \textbf{Choudhury et al.~[29]} \\
\hline
\textbf{Objective} & Behavioural mutation detection and evolution analysis &
 DNA-inspired bot detection using relative entropy & ML-based bot
 classification & Fake news and bot identification \\
\hline
\textbf{Techniques Used} & Bioinformatics-inspired alignment and mutation
 detection & DNA-sequence modelling with relative entropy & Supervised and
 unsupervised ML & Natural language processing \\
\hline
\textbf{Phenomena Studied} & Evolutionary changes in bot behaviour patterns &
 Behavioural regularity and deviations using entropy & Bot behaviour
 classification & Detection of manipulation patterns \\
\hline
\textbf{Validation} & Mutation trends across families; sequence analysis &
 Empirical entropy thresholds validated on labelled datasets & Accuracy,
 precision, and recall & Detection rates and dataset validation \\
\hline
\textbf{Results} & Evaluates inter-family and intra-family shared mutations &
 Identifies behavioural deviations and entropy-driven patterns & High-accuracy
 classification & Identifies bots with nuanced patterns \\
\hline
\end{tabular}
\end{table}

In contrast, our framework:
\begin{itemize}
    \item uses bioinformatics-inspired alignment to preserve positional
     information and shared structures across multiple sequences,
    \item explicitly categorises behavioural changes as insertions, deletions,
     substitutions, alterations and identity (match) events,
    \item analyses mutation trends within and across behaviourally defined
     families, rather than only at the account level,
    \item validates the approach by examining the consistency of mutation
     patterns with known family characteristics and by studying intra-family
     transfer of mutations.
\end{itemize} This moves from a purely detection-oriented perspective towards a
 richer, evolutionary account of how bots adapt over time.

\subsection{Limitations}

Two practical limitations of the family-modelling framework are important when
interpreting the results:
\begin{itemize}
    \item \textbf{Incomplete behavioural histories:} Sequences are built from
    posts collected at specific times rather than continuous streams. Deleted
    posts, protected accounts, or the 3{,}200-tweet retrieval limit can lead to
    partial views of early behaviour for long-lived bots.
    \item \textbf{Dependence on chosen features and encoding:} The methodology
    relies on seven categorical features and a specific encoding scheme.
    Incorporating additional information (e.g., interaction-network structure or
    richer content semantics) could further refine family boundaries.
\end{itemize}

At the same time, several aspects of the study provide \textbf{internal
robustness}. First, the dataset spans multiple platform generations and
contains millions of posts, which stabilises frequency-based estimates of
dominant blocks at the family level. Second, we restrict analyses to complete
calendar years and operate on derived behavioural features, reducing sensitivity
to transient content and to gaps in raw text availability. Third, the main
claims are supported by \emph{two complementary lenses} on behaviour change:
family discovery (via similarity over block-frequency vectors) and within-family
dynamics (via alignment and mutation classification). Consistency between these
two views strengthens confidence that the identified families and change
patterns are not artefacts of a single modelling choice.

The dynamic mutation framework introduces additional, largely methodological,
considerations:
\begin{itemize}
    \item \textbf{Alignment assumptions:} As with any MSA-based approach, the
    inferred alignment reflects the scoring scheme and gap penalties. In this
    work we use alignment primarily as a tool for \emph{localising} and
    \emph{counting} change, and the qualitative conclusions are driven by broad
    differences (e.g., deletion/substitution dominance and hotspot locations)
    rather than by any single aligned position.
    \item \textbf{Alteration threshold:} The 50\% similarity threshold separating
    alterations from insertions is interpretable and aligns with the intent of
    distinguishing ``modified'' versus ``new'' blocks. A small sensitivity check
    around this threshold (e.g., 40--60\%) would be a straightforward addition
    in future work and would help quantify the stability of fine-grained
    mutation counts.
\end{itemize}

Future work can therefore focus on \emph{incremental} extensions that preserve
interpretability: (i) lightweight sensitivity analyses for key thresholds and
similarity choices; (ii) incorporating a limited set of network or semantic
features to test whether families sharpen or split; and (iii) improving handling
of missingness by combining tweet-ID based rehydration, downsampling to equal
post budgets, or stratifying analyses by account longevity.

\section{Evaluating the Feasibility of Predicting Behavioural Evolution}

We now turn to RQ4 and ask whether the evolution-inspired framing can
support \emph{predictive} reasoning about bot behaviour. We evaluate this  by
integrating three components of our framework, namely family clustering,
mutation detection, and a new transferability layer, into a dynamic model of
behaviour change. The evaluation consists of two analyses and one case study:

\begin{itemize}
    \item Analysis~1: quantifying shared mutations within and between bot
     families.
    \item Analysis~2: evaluating mutation transferability between closely and
     distantly related bots.
    \item Case study: examining behavioural adaptations around trigger events
     (Christmas and Halloween).
\end{itemize}

Together, these components provide evidence about whether behavioural changes
propagate preferentially along family relationships and whether such
propagation can be used predictively.

\subsection{Analysis 1: Shared Mutations Within and Between Families}

\subsubsection*{Question and Objectives}

Analysis~1 asks whether there is a statistically significant difference in the
extent of shared mutations:
\begin{itemize}
    \item between bots within the same family, and
    \item between bots in different families.
\end{itemize} The objective is to quantify how strongly familial relationships
 shape the likelihood of shared mutations, using measures such as overlap,
 density and sparsity (Table~7.1).

\subsubsection*{Method}

Unlike the alignment-based analysis, Analysis~1 detects mutations \emph
{independently} within each bot sequence, without relying on pairwise alignment
or MSA. For each bot, we scan its behavioural sequence and classify blocks over
time as:

\begin{itemize}
    \item \textbf{Insertion:} a block that appears for the first time in the
     sequence.
    \item \textbf{Substitution:} a block differing in a single letter from a
     previously inserted block.
    \item \textbf{Identity:} a reoccurrence of a block already labelled as an
     insertion or substitution.
    \item \textbf{Deletion:} a previously present block that no longer appears
     after a given position.
\end{itemize}

The ``Identity'' category is excluded from shared-mutation calculations to focus on
primary changes. For each family, we record the set of mutation events for each
bot.

\paragraph{Within-family shared mutations.}

For any pair of bots $(bot_i, bot_j)$ within a family $C_f$, we define a shared
mutation as an identical mutation type (insertion, deletion, substitution)
occurring in the same block pattern. The shared-mutation score is the size of
the intersection of their mutation sets normalised by the larger of their
individual mutation counts (Equation~7.1). These scores form a sparse $n\times
n$ matrix $M$ for each family, where $M_{ij}$ is the proportion of shared
mutations between $bot_i$ and $bot_j$.

We summarise $M$ using:
\begin{itemize}
    \item average, median, minimum and maximum similarity,
    \item matrix sparsity: the proportion of zero entries in $M$ (Table~7.1).
\end{itemize}

\paragraph{Between-family shared mutations.}

To quantify shared mutations between families $C_a$ and $C_b$, we compute:
\begin{itemize}
    \item the total number of shared mutation events between the two families
     (Equation~7.2),
    \item a density measure: the proportion of shared mutations relative to the
     total number of mutations in both families (Equation~7.4).
\end{itemize} High Density ($\approx$ 1) relates to strong inter-family
 connections, with significant overlap in mutation-sharing behaviour, whilst
 Low Density ($\\<<$ 1) has weak inter-family connections, with minimal
 mutation-sharing behaviour.
\subsubsection*{Results}

Table~\ref{tab:sparse-matrix-results} summarises the within-family sparse
matrices. Key observations include:
\begin{itemize}
    \item \textbf{Family~3} (\emph{Content Multipliers}) is the densest family,
     with the lowest sparsity (about 14.7\%) and the highest average
     similarity ($\approx 0.38$), indicating strong mutation-sharing
     connections.
    \item \textbf{Family~4} (\emph{Informed Contributors}) shows moderate
     sparsity ($\approx 16\%$) and a relatively high average similarity
     ($\approx 0.36$), suggesting balanced diversity and connectivity.
    \item \textbf{Families~1 and 2} (\emph{Unique Tweeters} and \emph
     {Duplicators with URLs}) exhibit higher sparsity (around 30--39\%) and
     lower median similarities, indicative of more heterogeneous
     mutation-sharing patterns.
\end{itemize}
\begin{table}[htb!]
\centering
\caption{Summary of the Sparse Matrix Results for Each Bot Family}
\label{tab:sparse-matrix-results}
\begin{tabular}{lcccc}
\hline
\textbf{Metric} & \textbf{Family 1} & \textbf{Family 2} & \textbf
 {Family 3} & \textbf{Family 4} \\
\hline Average Similarity & 0.31 & 0.29 & 0.38 & 0.36 \\ Median Similarity  &
 0.13 & 0.14 & 0.30 & 0.26 \\ Max Similarity     & 1.0  & 1.0  & 1.0  & 1.0  \\
 Min Similarity     & 0.0  & 0.0  & 0.0  & 0.0  \\ Matrix Sparsity    & 0.30 &
 0.39 & 0.147 & 0.16 \\
\hline
\end{tabular}
\end{table}

Table~\ref{tab:shared-mutations-density} reports shared mutations and density
between families. All pairs have relatively low densities
(0.37--0.45), suggesting that most mutations remain family-specific. Among
these:
\begin{itemize}
    \item Families~3 and 4 share the most mutations (230) with a density of
     0.4440, indicating strong inter-family connectivity.
    \item Families~1 and 3, and Families~1 and 4, show moderate shared
     mutations (219 and 205, respectively) and densities around 0.44--0.45,
     suggesting that Family~1 acts as an intermediate connector.
    \item Families~2 and 3 exhibit the lowest shared mutations (167) and
     density (0.3719), indicating relatively distinct evolutionary paths.
\end{itemize}

\begin{table}[htb!]
\centering
\caption{Results of Shared Mutations and Density Test Between Families}
\label{tab:shared-mutations-density}
\begin{tabular}{lrrrr}
\hline
\textbf{Families} & \textbf{First Family Mutations} & \textbf{Second Family
 Mutations} & \textbf{Shared Mutations} & \textbf{Density} \\
\hline 1 and 2 & 221 & 167 & 154 & 0.3969 \\ 1 and 3 & 221 & 282 & 219 &
 0.4354 \\ 1 and 4 & 221 & 236 & 205 & 0.4486 \\ 2 and 3 & 167 & 282 & 167 &
 0.3719 \\ 2 and 4 & 167 & 236 & 161 & 0.3995 \\ 3 and 4 & 282 & 236 & 230 &
 0.4440 \\
\hline
\end{tabular}
\end{table}

\subsubsection*{Interpretation}

Analysis~1 supports three main points:
\begin{itemize}
    \item Bots within the same family share substantially more mutations than
     bots across families, especially in Families~3 and~4, which show dense
     intra-family connectivity.
    \item Inter-family densities are low, indicating that each family retains a
     substantial set of unique mutations and follows its own evolutionary
     trajectory.
    \item Family~3 appears as a central hub in mutation-sharing dynamics, with
     strong intra-family cohesion and substantial overlap with Family~4, while
     Families~1 and~2 show more uniform but weaker sharing.
\end{itemize}

These results provide direct evidence for RQ4 that familial structure is
meaningful for modelling where and how behavioural changes are likely to
propagate.

\subsection{Analysis 2: Mutation Transferability Within Families}

\subsubsection*{Question and Objectives}

Analysis~2 asks:
\begin{quote} To what extent can mutation transfers between bots be predicted
 based on the closeness of the relationship between source and target bots?
\end{quote}

The objectives are to:
\begin{enumerate}[leftmargin=*]
    \item analyse mutation transfer patterns within bot families,
    \item evaluate how closeness (cosine similarity) between bots influences the
     likelihood and accuracy of transfer,
    \item assess whether similarity can serve as a reliable predictor of
     mutation transfer.
\end{enumerate}

\subsubsection*{Method}

The analysis proceeds in five steps.

\paragraph{(1) Sampling most- and least-related bots.}

For each family, we compute the similarity matrix using cosine similarity
between behavioural sequences. We then:
\begin{itemize}
    \item rank bots by their average similarity to all other bots in the
     family,
    \item select the 10 most closely related bots (highest average similarity),
    \item select the 10 least closely related bots (lowest average similarity).
\end{itemize} All subsequent steps are applied separately to these two groups.

\paragraph{(2) Independent mutation detection.}

Mutations for each selected bot are detected independently as in Analysis~1,
again excluding the ``Identity'' category.

\paragraph{(3) Identifying mutation transfer events.}

A mutation transfer is defined as a mutation $\delta_k$ that:
\begin{itemize}
    \item first appears at position $k$ in a \emph{source} bot $S_i$,
    \item later appears in the same block pattern in a \emph{target} bot $S_j$
     at position $k$ or later in its sequence,
\end{itemize} with sequence position acting as a proxy for time
 (Section~5.3.4). Transfer requires matching both mutation type and block; an
 exception is made for cases where both bots reach the same block via insertion
 or substitution (e.g., one inserts a pattern, the other modifies an existing
 pattern to match it). The formal notation is given in Equation~7.5.

\paragraph{(4) Confusion matrices for transferability.}

For each pair of bots, we construct a confusion matrix over all possible
mutations in the family (Table~7.5):

\begin{itemize}
    \item TP: mutations shown by both source and target,
    \item FP: mutations shown by target but not source,
    \item FN: mutations shown by source but not target,
    \item TN: mutations not shown by either bot.
\end{itemize}

\paragraph{(5) Family-level aggregation.}

We aggregate TP, FP, FN and TN across all pairs in a family and compute:
\[
\text{Precision}_{family} = \frac{\text{TP}}{\text{TP} + \text{FP}},\quad
\text{Recall}_{family} = \frac{\text{TP}}{\text{TP} + \text{FN}},\quad
\text{F1}_{family} = \frac{2 \cdot \text{Precision}_{family} \cdot \text
 {Recall}_{family}}{\text{Precision}_{family} + \text{Recall}_{family}}.
\] Table~\ref{tab:agg-metrics-related-bots} reports these metrics for each
 family and each group (most- and least-related bots). 
\begin{table}[ht]
\centering
\caption{Aggregated metrics and scores for most 10 and least 10 related bots
 across bot families}
\label{tab:agg-metrics-related-bots}
\begin{tabular}{llrrrrrrr}
\hline
\textbf{Family} & \textbf{Samples Group} & \textbf{TP} & \textbf{FP} & \textbf
 {FN} & \textbf{TN} & \textbf{Precision} & \textbf{Recall} & \textbf
 {F1 Score} \\
\hline 1 & 10 most related bots  & 10631   & 3891   & 6626    & 29766   &
 0.7321 & 0.6160 & 0.6691 \\ 1 & 10 least related bots & 18388   & 12815  &
 9540    & 41616   & 0.5893 & 0.6584 & 0.6219 \\ 2 & 10 most related bots  &
 1540    & 247    & 227     & 3059    & 0.8618 & 0.8715 & 0.8666 \\ 2 & 10
 least related bots & 166688  & 41772  & 163839  & 391124  & 0.7996 & 0.5043 &
 0.6185 \\ 3 & 10 most related bots  & 2258848 & 114128 & 8320    & 3763600 &
 0.9519 & 0.9963 & 0.9736 \\ 3 & 10 least related bots & 14937   & 30537  &
 12302   & 106785  & 0.3285 & 0.5484 & 0.4108 \\ 4 & 10 most related bots  &
 317557  & 131387 & 60689   & 644229  & 0.7073 & 0.8396 & 0.7678 \\ 4 & 10
 least related bots & 144992  & 45198  & 99374   & 377910  & 0.7624 & 0.5933 &
 0.6673 \\
\hline
\end{tabular}
\end{table}

\begin{figure}
    \centering
    \includegraphics[width=1\linewidth]{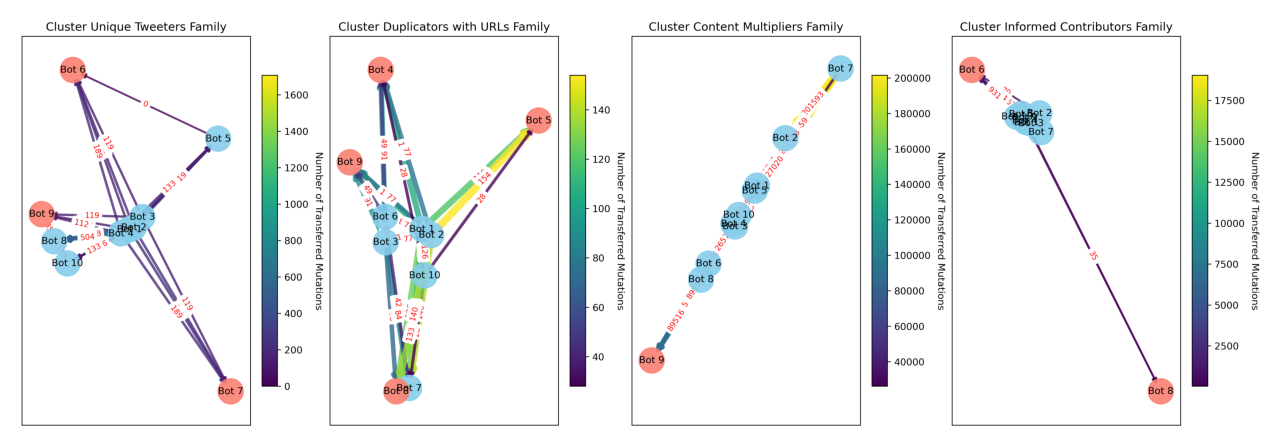}
    \caption{Network visualization of transferring mutations between the 10 most
     closely related bots within each family}
    \label{fig:7.3}
\end{figure}

Figure~\ref{fig:7.3} visualises transfer networks among the 10 most closely
related bots per family.
\subsubsection*{Results}

Table~\ref{tab:agg-metrics-related-bots} shows that, across all four families,
the 10 most closely related bots consistently achieve higher F1 scores than the
10 least related bots. In particular:

\begin{itemize}
    \item For Family~3 (\emph{Content Multipliers}), the most-related group
     attains very high precision ($\approx 0.95$), recall ($\approx 0.996$) and
     F1 ($\approx 0.97$), while the least-related group drops to precision
     $\approx 0.33$ and F1 $\approx 0.41$.
    \item For Families~1, 2 and 4, F1 for the most-related group ranges from
     about 0.67 to 0.87, while the least-related group achieves F1 between
     about 0.62 and 0.67.
    \item The most-related groups typically achieve higher precision, indicating
     few false positives when predicting that a mutation transfers from a
     source to a similar target.
\end{itemize}

The network visualisations in Figure~\ref{fig:7.3} further illustrate these
patterns. For the 10 most closely related bots:
\begin{itemize}
    \item Family~3 shows dense, reciprocal transfer links among many bots, with
     one bot (Bot~9) acting mainly as a receiver.
    \item Family~1 shows sparser connections, with a few bots (e.g., Bots~6, 7,
     9) acting as central hubs of mutation transfer.
    \item Families~2 and~4 show intermediate connectivity.
\end{itemize}

\subsubsection*{Interpretation}

Analysis~2 provides strong evidence that:
\begin{itemize}
    \item mutation transfers are far more predictable between \emph
     {similar} bots than between dissimilar ones,
    \item Family~3 has particularly strong, cohesive transfer dynamics, but is
     also sensitive to the choice of subgroups: performance drops sharply when
     distant bots are considered,
    \item families differ in how tightly similarity constrains transfer,
     suggesting that sub-family structure (especially in Family~3) may further
     improve predictive accuracy.
\end{itemize}

These findings support the claim that the evolution-inspired framing can be
leveraged predictively: given a mutation in one bot, we can estimate with
reasonable precision which similar bots are likely to exhibit the same change.

\subsection{Case Study: Trigger Events and Emoji Usage}

\subsubsection*{Question and Objectives}

The case study investigates whether bot behaviour around specific trigger
events (Christmas and Halloween) can be analysed to evaluate the feasibility of
predicting future adaptations, and how closely related bots contribute to these
changes.

The objectives are to:
\begin{itemize}
    \item examine how bots adapt their behaviour before, during and after
     events,
    \item compare responses across families,
    \item assess whether family membership and closeness help explain
     behavioural propagation.
\end{itemize}

\subsubsection*{Method}

The case study uses the raw bot data, focusing on emoji usage as a clear and
event-specific behavioural feature.

\paragraph{Trigger events and emoji sets.}

Two global events are considered:
\begin{itemize}
    \item Christmas (25 December),
    \item Halloween (31 October).
\end{itemize} Using Emojipedia, the ten most event-specific emojis for each
 event are identified (e.g., Santa, gifts, snowmen, trees for Christmas;
 pumpkins, ghosts, bats, skulls for Halloween).

\paragraph{Temporal windows.}

For each bot and each event, emoji usage is monitored:
\begin{itemize}
    \item five days \emph{before} the event,
    \item on the \emph{event day},
    \item five days \emph{after} the event.
\end{itemize} Monthly usage profiles across the year are also computed to locate
 seasonal peaks (Figures~7.4 and~7.5).
\begin{figure}
    \centering
    \includegraphics[width=0.9\linewidth]{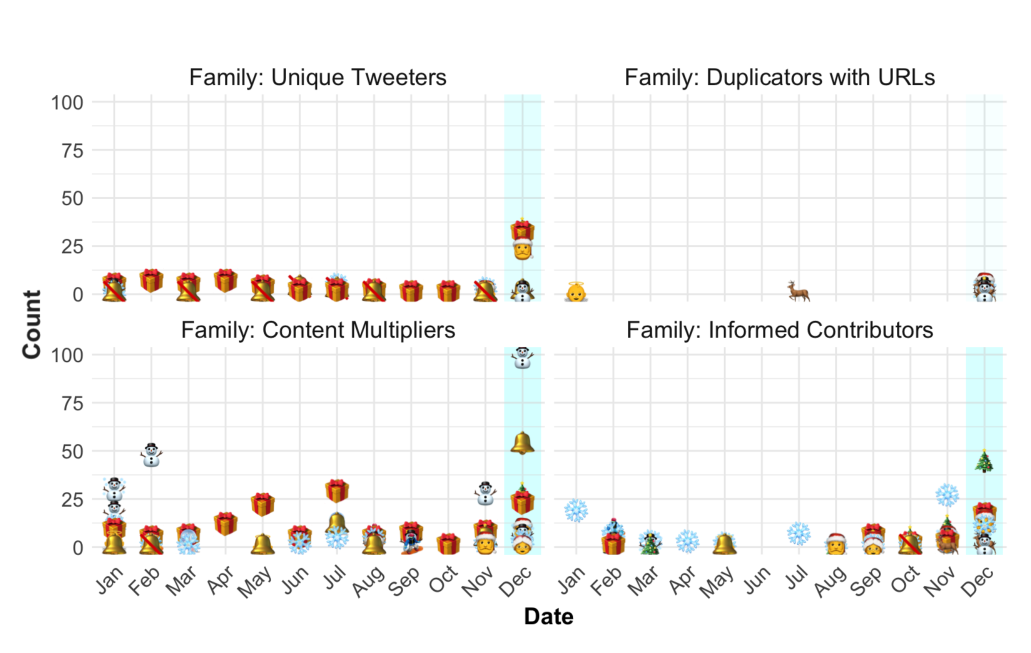}
    \caption{Bot families’ behavioural changes in response to the Christmas
     event, represented by the usage of Christmas-related emojis.}
    \label{fig:7.4}
\end{figure}
\begin{figure}
    \centering
    \includegraphics[width=0.9\linewidth]{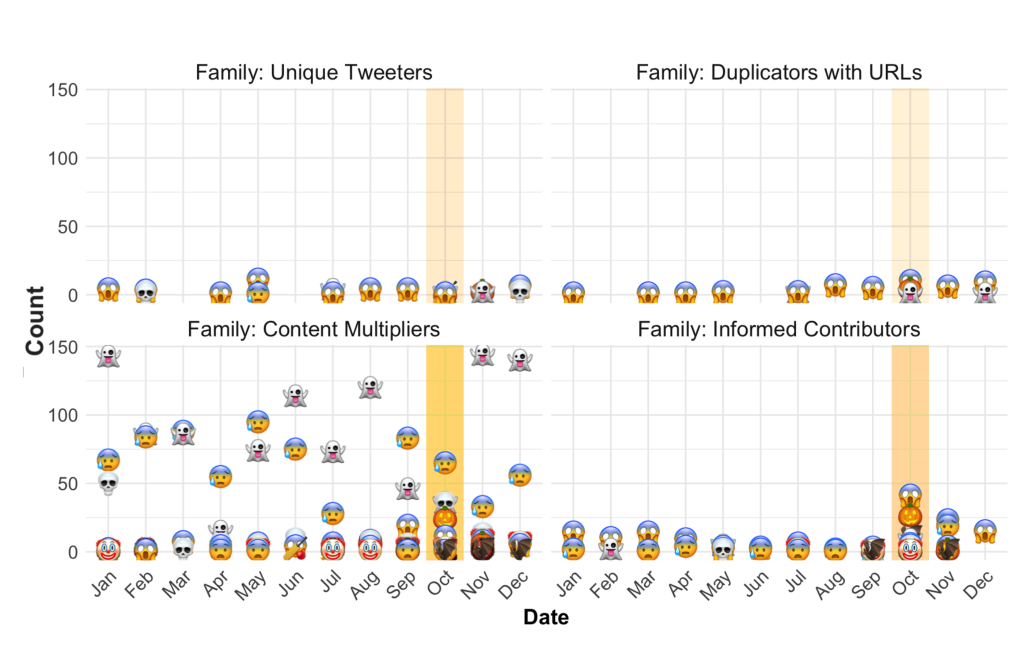}
    \caption{Bot families' behavioural changes in relation to Halloween}
    \label{fig:7.5}
\end{figure}
\paragraph{Family comparison and participation.}

For each family:
\begin{itemize}
    \item we compute the total volume of event-specific emoji usage across
     months (Figures~7.4, 7.5),
    \item we count how many bots use at least one event-specific emoji before,
     during and after the event (Figures~7.8, 7.9),
    \item we compute the mean similarity (as in earlier sections) among bots
     that participate in emoji changes (Table~\ref{tab:emoji-trigger-events}).
\end{itemize}

\paragraph{Sequential adaptation and backward reasoning.}

Within Family~3, we visualise emoji usage around the events at the level of
individual bots using density ridges (Figures~7.6, 7.7), showing how usage
peaks and propagates over days. We then relate these patterns back to the
events to reason about causal triggers and propagation paths.
\begin{figure}
    \centering
    \includegraphics[width=0.9\linewidth]{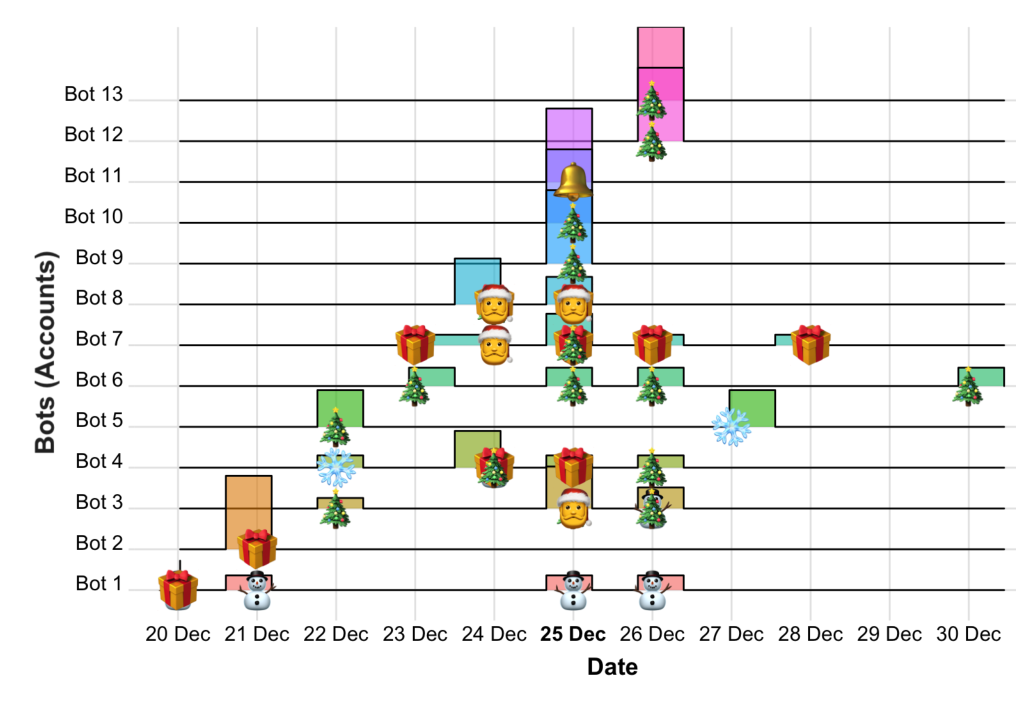}
    \caption{Gradual adaptation of emoji usage by bots around Christmas Day}
    \label{fig:7.6}
\end{figure}
\begin{figure}
    \centering
    \includegraphics[width=0.9\linewidth]{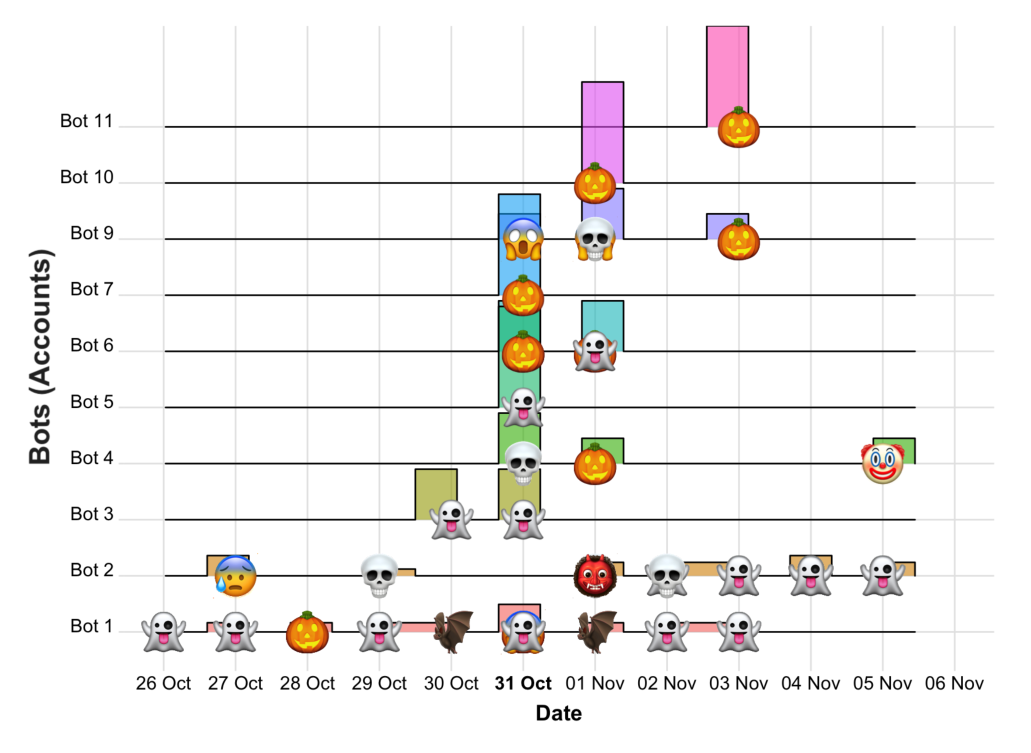}
    \caption{Adaptation of emoji usage around  Halloween}
    \label{fig:7.7}
\end{figure}
\subsubsection*{Results}

\paragraph{Different families, different responses.}

Figures~\ref{fig:7.4} (Christmas) and~\ref{fig:7.5} (Halloween) show aggregated
emoji usage per month for each family. In both cases:
\begin{itemize}
    \item usage peaks in December for Christmas emojis and in October for
     Halloween emojis,
    \item Family~3 (\emph{Content Multipliers}) has by far the highest usage,
     reflecting their engagement-oriented strategy,
    \item Family~4 (\emph{Informed Contributors}) shows moderate emoji usage,
    \item Family~1 (\emph{Unique Tweeters}) uses emojis sparingly,
    \item Family~2 (\emph{Duplicators with URLs}) shows almost no event-specific
     emoji usage.
\end{itemize}

\paragraph{Gradual adaptation around events.}

Figures~\ref{fig:7.6} and~\ref{fig:7.7} focus on Family~3 and plot emoji usage
around Christmas and Halloween, respectively. They reveal that:
\begin{itemize}
    \item emoji usage peaks sharply on the event day (25 December, 31 October),
     with a secondary peak on the following day,
    \item adoption is \emph{sequential}: some bots start using event-specific
     emojis days before the event (e.g., Bot~1), followed by others
     (Bot~2, Bot~3, and so on),
    \item usage then tapers off gradually, rather than disappearing
     immediately.
\end{itemize} This pattern is consistent with behaviour changes propagating
 within the family over time, rather than all bots reacting independently and
 simultaneously.

\paragraph{Participation rates.}
\begin{figure}
\centering
\begin{subfigure}{.5\textwidth}
  \centering
  \includegraphics[width=.4\linewidth]{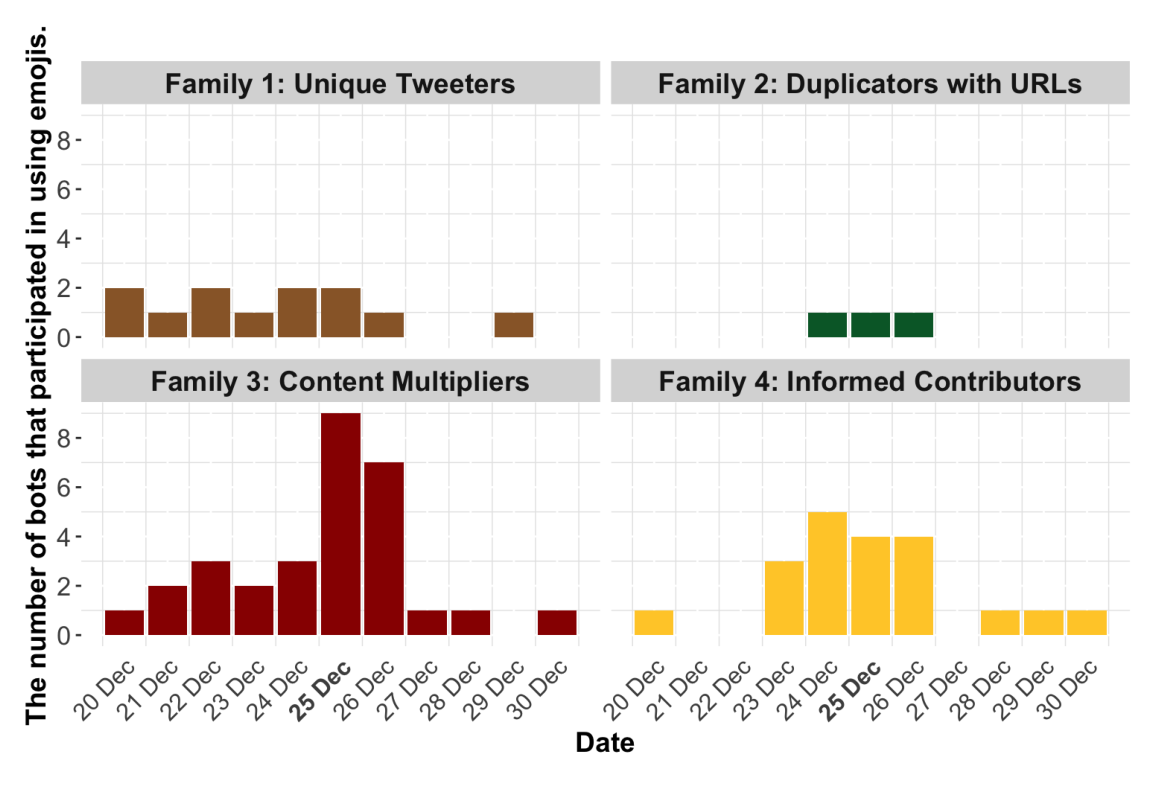}
  \caption{Christmas}
  \label{fig:Christmas}
\end{subfigure}%
\begin{subfigure}{.5\textwidth}
  \centering
  \includegraphics[width=.4\linewidth]{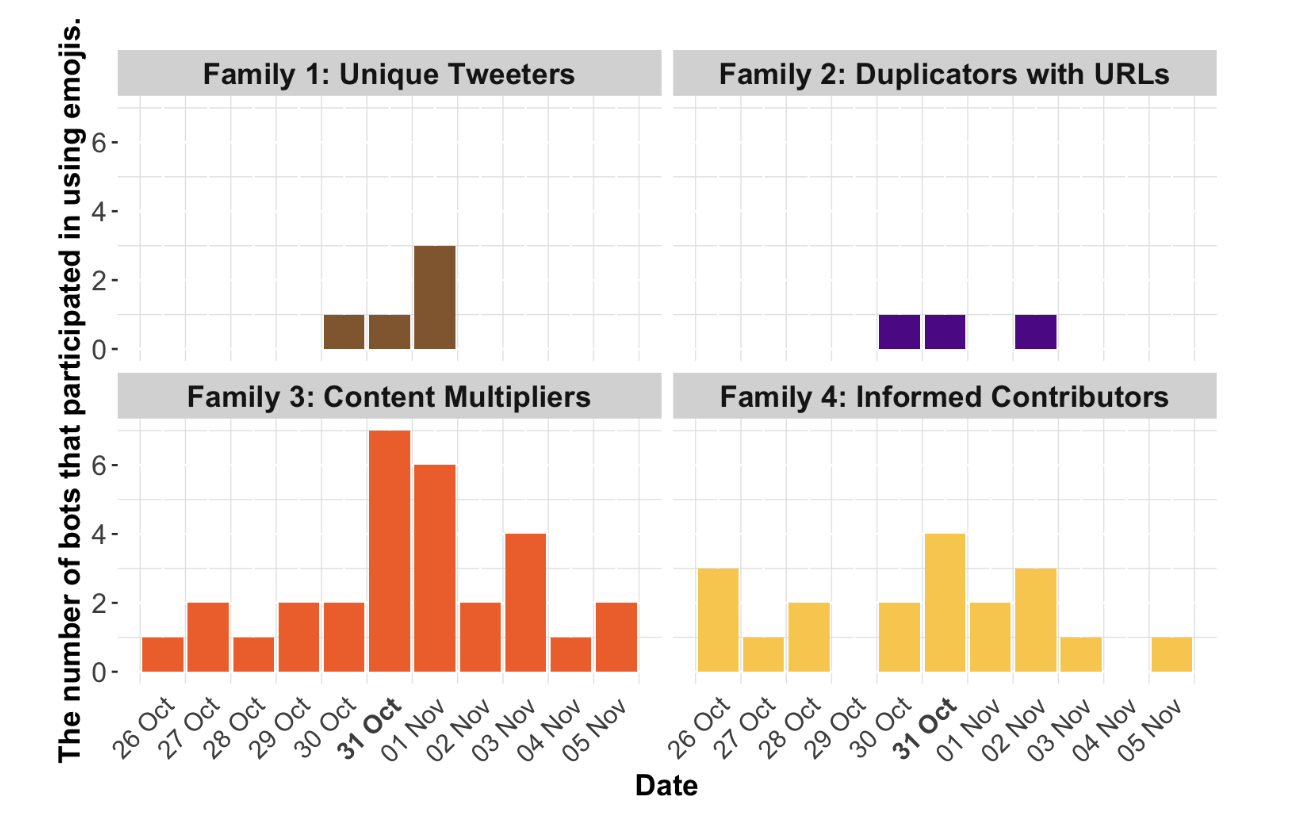}
  \caption{Halloween}
  \label{fig:Halloween}
\end{subfigure}
\caption{Count of bots using event-specefic emojis before, during, and after the
 events}
\label{fig:bot-participation}
\end{figure} Figure~\ref{fig:bot-participation} show how many bots in each
 family use event-specific emojis before, during and after the events. In both
 cases:
\begin{itemize}
    \item participation increases towards the event, peaks on the event day, and
     decreases afterwards,
    \item Family~3 consistently has the largest number of participating bots,
    \item Family~4 shows moderate participation,
    \item Families~1 and~2 have minimal participation.
\end{itemize}

\paragraph{Behavioural relationships among participating bots.}

Table~\ref{tab:emoji-trigger-events} reports, for each event and family:
\begin{itemize}
    \item the number of bots participating in emoji-related changes,
    \item the mean behavioural similarity among those bots.
\end{itemize} It reveals that:
\begin{itemize}
    \item Family~3 has the largest number of participating bots (13 for
     Christmas, 10 for Halloween), but the \emph{lowest} mean similarity
     (0.16 and 0.21),
    \item Families~1 and~4 have fewer participating bots (5--8) but \emph
     {higher} mean similarity (around 0.46--0.57 for Family~1 and 0.49--0.51
     for Family~4),
    \item Family~2 has only one participating bot in each event and thus no
     defined similarity.
\end{itemize}
\begin{table}[htb!]
\centering
\caption{Average behavioural similarity and number of bots participating in
 emoji changes during trigger events}
\label{tab:emoji-trigger-events}
\begin{tabular}{llcccc}
\hline
\textbf{Event} & \textbf{Metric} & \textbf{Family 1} & \textbf
 {Family 2} & \textbf{Family 3} & \textbf{Family 4} \\
\hline Christmas & Number of Bots   & 5  & 1 & 13 & 8  \\ Christmas & Similarity
 Mean  & 0.46 & 0 & 0.16 & 0.49 \\ Halloween & Number of Bots   & 5  & 1 & 10 &
 6  \\ Halloween & Similarity Mean  & 0.57 & 0 & 0.21 & 0.51 \\
\hline
\end{tabular}
\end{table}

\subsubsection*{Interpretation}

The case study yields four main insights:

\begin{itemize}
    \item \textbf{Event-driven peaks.} Bot behaviour is clearly shaped by
     external temporal triggers: event-specific emoji usage peaks exactly on
     Christmas Day and Halloween, confirming that our sequence encoding
     captures meaningful, time-sensitive behaviours.
    \item \textbf{Family-specific adaptation styles.} Families differ
     systematically:
    \begin{itemize}
        \item Family~3 is highly active and responsive, using event-specific
         emojis extensively.
        \item Family~4 shows moderate, coordinated adaptation.
        \item Families~1 and~2 show limited, niche responses.
    \end{itemize} These patterns align with earlier characterisations:
     engagement-heavy behaviour in Family~3 and URL-centric promotion in
     Family~2.
    \item \textbf{Sequential propagation.} The ridge plots for Family~3
     (Figures~7.6, 7.7) show that emoji usage emerges in a subset of bots and
     then spreads, supporting the idea of change propagation within families
     rather than purely independent reactions.
    \item \textbf{Closeness vs.\ response.} The relationship between similarity
     and adaptation depends on how typical the behaviour is:
    \begin{itemize}
        \item In families where emojis are \emph{uncommon} (Families~1 and~2),
         participating bots tend to be more similar, suggesting that novel
         behaviours are first explored by closely related bots.
        \item In Family~3, where emoji usage is routine, the very high
         participation and low similarity suggest that adaptation is diffuse
         and not restricted to the most similar bots.
    \end{itemize}
\end{itemize}

Overall, the case study demonstrates that event-driven behavioural changes can
be detected and interpreted within our framework, and that family structure
helps explain how these changes propagate.

\subsection{Implications for Predictive Modelling (RQ4)}

Taken together, Analysis~1, Analysis~2 and the case study address RQ4: they show
that the evolution-inspired framing is not merely descriptive but can underpin
predictive reasoning about bot behaviour.

\begin{itemize}
    \item \textbf{Family-level structure} (Analysis~1) establishes that
     mutations are more likely to be shared within families than across
     families, and that some families (notably Family~3) act as hubs of
     mutation sharing.
    \item \textbf{Similarity-based transferability} (Analysis~2) demonstrates
     that, within a family, the closeness of bots is a strong predictor of
     which bots will share mutations, with very high precision and F1 for the
     most similar bots.
    \item \textbf{Event-driven adaptation} (case study) shows that families
     respond differently to external triggers and that behaviour changes
     propagate sequentially within families, in ways that can be related to
     family characteristics and similarity relationships.
\end{itemize}

These three evaluation components link to specific outcomes: increased
likelihood of mutation sharing within families, higher propagation rates
between similar bots, and family differences in adaptation to trigger events.
Collectively, they provide evidence that the evolution-inspired framing is
useful for anticipating where, when and how promotional bots will adapt their
behaviour.

\section{Conclusion}

By encoding multi-feature post behaviour into unified sequences and applying
hierarchical clustering, we demonstrate that promotional Twitter bots can be
organised into four coherent behavioural families and that members of each
family exhibit similar patterns of behavioural evolution. The combination of
shared core blocks, family-specific unique patterns, and family-consistent
temporal trajectories offers a rich basis for both descriptive understanding
and predictive modelling of bot behaviour.

Extending this with a sequence-analysis mutation framework, we show that
changes in bot behaviour can be modelled as additions, removals and
modifications of activity patterns detected via multiple sequence alignment and
mutation classification. Deletions and substitutions dominate, insertions are
rare, and mutation hotspots and substitution patterns differ systematically
across the four families.

Finally, the evaluation phase demonstrates that this evolution-inspired framing
has predictive value. Mutations are more likely to be shared within families
than across them; closely related bots are more likely to share and propagate
mutations than distant ones; and responses to trigger events such as Christmas
and Halloween show family-specific, partially predictable patterns. Together,
these results support sequence-based family modelling of promotional bots
coupled with a detailed account of their behavioural dynamics. This, in turn,
points towards more adaptive and robust detection approaches that exploit not
only static behavioural signatures but also characteristic patterns of
behavioural change over time and across related accounts.

\section*{Availability of data and materials}
This study uses bot account identifiers drawn from publicly available research
corpora (including datasets associated with the MIB project and Botometer's Bot
Repository) and derives behavioural sequences and aggregate statistics from the
corresponding tweet collections. Because redistribution of full tweet content
can be constrained by platform terms, we focus on shareable derived artefacts
that preserve scientific utility without exposing raw content. Specifically, we
will make available (i) the feature-extraction and sequence-construction code,
(ii) the behavioural block vocabulary and per-account sequences in encoded form,
(iii) similarity matrices and clustering assignments, and (iv) aggregate
mutation statistics and trigger-event summaries sufficient to reproduce the
figures and tables in this paper. Where appropriate, tweet identifiers can be
provided to support rehydration-based replication.

\section*{Code availability}
All analysis code required to reproduce the sequence construction, family
clustering, alignment, and mutation analyses will be released in a public
repository at the time of publication.

\section*{Competing interests}
The authors declare that they have no competing interests.

\bibliographystyle{abbrvnat}
\bibliography{botrefs}
\end{document}